\documentclass[aps,prd,showpacs,a4paper,%
  superscriptaddress,preprintnumbers,10pt]{revtex4}
\usepackage{epsfig,graphicx,bm}
\begin{document}
\newcommand{\nd}{\noindent}
\newcommand{\beq}{\begin{equation}}
\newcommand{\eeq}{\end{equation}}
\newcommand{\barr}{\begin{eqnarray}}
\newcommand{\earr}{\end{eqnarray}}
\newcommand{\ba}{\begin{array}}
\newcommand{\ea}{\end{array}}
\newcommand{\bfk}{\mbox{\boldmath $k$}}
\title{Glueball production in radiative $\bm{J/\psi,\Upsilon}$ decays}
\author{Maurizio Melis}\email{maurizio.melis@ca.infn.it}
\altaffiliation[present address: ]{Dipartimento di Matematica,
 Universit\`a di Cagliari, viale Merello 92, 09123 Cagliari, Italy}
\affiliation{Istituto Nazionale di Fisica Nucleare,
 Sezione di Cagliari, and \\
 Dipartimento di Fisica, Universit\`a di Cagliari \\
 Casella Postale n. 170, I-09042 Monserrato (CA), Italy}
\author{Francesco Murgia}\email{francesco.murgia@ca.infn.it}
\affiliation{Istituto Nazionale di Fisica Nucleare,
 Sezione di Cagliari, and \\
 Dipartimento di Fisica, Universit\`a di Cagliari \\
 Casella Postale n. 170, I-09042 Monserrato (CA), Italy}
\author{Joseph Parisi$\,$ }\email{Joseph.Parisi@th.u-psud.fr}
\affiliation{ Laboratoire de Physique Th\'eorique,
 Universit\'e de Paris-Sud\\ 
 Laboratoire associ\'e au Centre Nationale de la
 Recherche Scientifique UMR 8627,\\
 B\^atiment 210, 91405 Orsay Cedex France}
\vspace{8pt}

\date{\today}

\begin{abstract}
Using a bound-state model of weakly bound gluons for glueballs made
of two gluons and a natural generalization of the perturbative QCD formalism
for exclusive hadronic processes, we present results for glueball production
in radiative $J/\psi,\Upsilon$ decays into several possible glueball states,
including $L\not= 0$ ones. We perform a detailed phenomenological analysis,
presenting results for the more favored experimental candidates and for
decay angular distributions.

\end{abstract}
\maketitle
\section{Introduction}
\label{intr}

The unambiguous experimental observation of two- or three-gluon bound states
should be an important test of Quantum Chromodynamics and of its internal
consistence. In the last years a huge experimental and theoretical work
was in fact dedicated to this goal. Unfortunately, lower mass glueball
states are supposed to fall in a mass range (a few GeV) which is already
largely occupied by mesonic and/or hadronic states, including exotic states,
hybrids etc. Moreover, glueball states are probably not narrow enough
to allow a very clear observation.
However, the intense experimental search \cite{exp,zou}, complemented by
lattice QCD calculations \cite{morn,wein} and several other theoretical contributions, like bag models \cite{carl}, flux-tube models \cite{isgu},
QCD sum rules \cite{nari},
has given evidence to a few possible candidates for two-gluon bound states.
Typically, experimental searches for glueball candidates have been performed
in favorable experimental environments. A good place to search for
glueballs is in radiative $J/\psi,\Upsilon$ decays,
because of the ``gluon rich environment'' produced in the $Q\bar{Q}$
decay, $Q\bar{Q}\to\gamma g g$, which should be particularly suitable for
subsequent glueball formation.
Two other processes of interest are  central production in hadronic collisions
and $\bar{p} p$ annihilation in the few GeV energy range. 
It is interesting to observe that one can take advantage also of
negative-answer experiments, like e.g. hadron production in photon-photon
collisions. In this case, meson production is expected to be favored 
with respect to glueball formation, which should only proceed via
higher order (in the e.m. coupling constant) contributions.
In some sense, the situation is reversed compared to radiative quarkonium
decays. From a more quantitative point of view,
this implies that the so-called ``stickiness'' parameter \cite{chan},
${\cal S}=[\Gamma(J/\psi,\Upsilon\to\gamma\,{\cal R})\,
PS(\gamma\gamma\to {\cal R})]/[\Gamma(\gamma\gamma\to {\cal R})\,
PS(J/\psi,\Upsilon\to\gamma\,{\cal R})]$, where ${\cal R}$ is a hadronic resonance
and $PS$ denotes the phase space factor, should be  much greater for
glueballs as compared to ordinary mesonic states.
The present status of the glueball search program can be summarized by
recalling that there are two main candidates for glueball states
\cite{exp,zou,pdg}:
a scalar, $0^{++}$ state with mass $M_{0^{++}} \simeq 1.5 $ or 1.7 GeV;
a tensor, $2^{++}$ (or $4^{++}$) state with mass
$M_{J^{++}}\simeq 2.22$ GeV; these observations are supported
also by lattice QCD calculations \cite{morn,wein}.
Other possible experimental candidates are \cite{zou}:
a pseudoscalar, $0^{-+}$ state with mass $M_{0^{-+}}=2.14$ GeV;
a vector, $1^{++}$ state with mass $M_{1^{++}}= 2.34$ GeV;
a pseudotensor, $2^{-+}$ state with mass $M_{2^{-+}}=2.04$ GeV;
a $3^{++}$ state with mass $M_{3^{++}}=2.0$ GeV.
Present lattice calculations, however, predict larger masses
for most of these states \cite{morn}.
Other glueball candidates have also been proposed.
In the following, we will not consider these additional states,
limiting ourself to discuss some selected examples in detail;
it will then be clear how to deal with the remainder.
Additional information can be found in ref.s \cite{exp,zou} and
references therein.

Motivated by the fundamental role played by $J/\psi,\Upsilon$ radiative
decays in the field of glueball search, in this paper we study in detail
these decay processes and analyze their implications for glueball production.

To this end, we use an approach previously developed by some of us and
already applied to several processes involving the production and/or decay
of hadronic states (including $q\bar{q}$ mesons and two-gluon bound
states) at high energies \cite{Kada,Houra,Kessl,Murgia}.
This approach is a generalization of the formalism developed,
starting from the early 80's, in the framework of perturbative QCD
for the study of exclusive hadronic processes at
high transfer momentum \cite{Brod}.
Our model assumes that the dominant contribution to the production/decay
of a hadronic state at large energy scale is given by its leading Fock state,
considered as a weakly-bound state of valence constituents,
with any internal angular orbital momentum.
It has also been shown that this approach reduces to the usual perturbative
QCD results in the case of $L=0$ states.
Let us clarify from the start that, being based on perturbative QCD,
this model is not suitable for predicting physical observables
which are nonperturbative in their essence.
Thus, for example, it cannot give sensible predictions for the
masses and decay constants of light $q\bar{q}$ and two-gluon states.
Rather, these quantities must be taken as input parameters of our
model, obtained from experimental results or other nonperturbative approaches,
like e.g. lattice QCD calculations or QCD sum rules.
In some cases, like for decay-product angular distributions,
it is possible to give predictions for physical observables which
have little dependence on these nonperturbative inputs, as will be shown
in the following.

Let us remind that our approach was already applied to
glueball production in radiative quarkonium decays in Ref.\cite{Kada},
in the case of a nonrelativistic distribution amplitude for the two-gluon
bound state.
Here we extend and generalize the analysis to the case of a generic
glueball distribution amplitude; we also present analytical and
numerical results obtained using several different models for the
glueball distribution amplitude. 

We derive and discuss, in the framework of our model, a number
of general analytical results for scattering amplitudes, angular
distributions etc., that may be useful independently of the numerical
results presented here. This should hopefully facilitate the use of our
approach for future, more detailed investigations along this lines.

The paper is organized as follows: in section II we summarize
our approach and present all relevant relations required.
In section III we consider in detail the radiative decay
$J/\psi,\Upsilon\to \gamma\, G$, deriving the expressions of the
helicity scattering amplitudes for several possible two-gluon bound states
with $S=0,1,2$, $L=0,1,2,3,4$ and $J=0,1,2,3,4$;
we also derive the expression of the decay photon angular distribution
in the quarkonium rest frame.
Section IV is devoted to the presentation and discussion
of several numerical results and predictions; we give our final
comments and conclusions in section V.
A few appendices contain most of the analytical
results for the helicity scattering amplitudes and some useful
analytical integrals required in the calculations. 

\section{\label{genris}
Description of the formalism and general results\protect\\
for helicity scattering amplitudes}

In this Section we outline the formalism utilized in the following
calculations and present some general expressions for the scattering 
amplitudes.
We limit ourselves to summarize the main steps of our approach.
The interested reader will find more details in Ref.s
\cite{Kada,Houra,Kessl,Murgia}, where the same approach was formulated
and applied to several hadronic production/decay processes
at high energy scales.

The basic ingredients we need in order to evaluate observables for the
process $J/\psi, \Upsilon\to\gamma\, G$ are the corresponding helicity
scattering amplitudes, $A_{\lambda_\gamma,\lambda_G;\lambda_Q}$,
where the $\lambda$'s indicate the helicities of the corresponding particles.
{}From now on we always indicate by the subscript ${\cal Q}$
all quantities related to the generic $Q\bar{Q}$ bound state,
${\cal Q}=J/\psi, \Upsilon$.

Notice that we know from first principles that the helicity amplitudes
$A_{\lambda_\gamma,\lambda_G;\lambda_{{\cal Q}}}$
can be written, in the quarkonium rest frame, as

\begin{equation}
A_{\lambda_\gamma,\lambda_G;\lambda_{{\cal Q}}}
(\theta_\gamma,\phi_\gamma) =
\hat{A}_{\lambda_\gamma,\lambda_G}\ e^{i\lambda_{{\cal Q}}\phi_\gamma}
\ d^1_{\lambda_{{\cal Q}},\lambda_\gamma-\lambda_G}(\theta_\gamma)\, ,
\label{ared}
\end{equation}

\noindent where $\theta_\gamma$, $\phi_\gamma$ are respectively the
polar and azimuthal angles specifying the photon outgoing direction
in the quarkonium rest frame, $d^{j}_{\lambda,\lambda'}$
are the usual Wigner rotation matrices
and $\hat{A}_{\lambda_\gamma,\lambda_G}$
are ``reduced'' amplitudes, depending only on
$\lambda_\gamma$, $\lambda_G$ and on the dynamics of the decay process.
Notice also that, since $d^j_{\lambda,\lambda'}(\theta\to0) =
\delta_{\lambda,\lambda'}$, Eq.~(\ref{ared}) gives

\begin{equation}
A_{\lambda_\gamma,\lambda_G;\lambda_{{\cal Q}}}(\theta_\gamma=\phi_\gamma=0) =
\hat{A}_{\lambda_\gamma,\lambda_G}\ 
\delta_{\lambda_{{\cal Q}},\lambda_\gamma-\lambda_G}\,.
\label{ared0}
\end{equation}

This means that we can evaluate the helicity amplitudes
$A_{\lambda_\gamma,\lambda_G;\lambda_{{\cal Q}}}$ in the simple kinematical
configuration $\theta_\gamma=\phi_\gamma=0$, find the reduced amplitudes
$\hat{A}_{\lambda_\gamma,\lambda_G}$ from Eq.~(\ref{ared0}) and utilize
Eq.~(\ref{ared}) to reconstruct the full amplitudes.

In the framework of perturbative QCD and its factorization theorems
for exclusive hadronic processes at high energy scales \cite{Brod}
(in our case, the ${Q}\bar{Q}$ bound-state mass, $M_{{\cal Q}}$),
the helicity amplitudes $A_{\lambda_\gamma,\lambda_G;\lambda_{{\cal Q}}}$
for the physical process 
may be evaluated starting from the helicity amplitudes for
the elementary hard process ${Q}\bar{Q}\to \gamma g^* g^*$,
$T_{\lambda_\gamma,\lambda_1,\lambda_2;\lambda_Q,\lambda_{\bar{Q}}}$,
where $\lambda_1,\lambda_2$ are the helicities of the two
final (virtual) gluons.

The partonic  process is schematically represented in Fig.~\ref{kin},
both in the $Q\bar{Q}$ c.m. ref. frame (the quarkonium rest frame) (A)
and in the glueball rest frame (B), where the two constituent gluons
are moving with relative momentum $2\bm{k}$.
As a first step, using the well-known nonrelativistic
color-singlet model for the quarkonium bound state, we can easily find
the connection between the helicity amplitudes
$T_{\lambda_\gamma,\lambda_1,\lambda_2;\lambda_Q,\lambda_{\bar{Q}}}$
and those relative to the process ${\cal Q}\to\gamma g^* g^*$,
$M_{\lambda_\gamma,\lambda_1,\lambda_2;\lambda_{{\cal Q}}}$.
A further step will in turn connect the amplitudes $M$ to the hadronic
amplitudes $A$, Eq.~(\ref{ared}),
which describe the full hadronic process.

Let us remark at this point that in the last years the crucial role
of color-octet contributions in hadronic processes involving the
production/decay of charmonium bound states has been discussed in the
framework of nonrelativistic QCD (NRQCD) \cite{nrqcd}.
It has been shown that color-octet contributions are crucial
for the consistence of a nonrelativistic model of quarkonium.
To our knowledge, only few attempts have been made to apply NRQCD and
color-octet contributions to the treatment of exclusive charmonium decays
\cite{kroll}. These attempts have shown that color-octet contributions
are probably of little relevance in the case of $J/\psi$, $\Upsilon$
exclusive decays.
In what follows, then, we stick to the usual color-singlet model
for the $^3S_1$, $Q\bar{Q}$ bound state.
In this case we can write:

\widetext

\begin{equation}
M_{\lambda_\gamma,\lambda_1,\lambda_2;\lambda_{{\cal Q}}}
(\theta_\gamma,\phi_\gamma;\beta,\theta,\phi)=
\frac{f_{{\cal Q}}}{2\sqrt{3}}\,\sum_{\lambda_Q,\lambda_{\bar{Q}}}
C^{1/2,\ 1/2,\ 1}_{\lambda_Q,\ \lambda_{\bar{Q}},\ \lambda_{{\cal Q}}}
\,T_{\lambda_\gamma,\lambda_1,\lambda_2;\lambda_Q,\lambda_{\bar{Q}}}
(\theta_\gamma,\phi_\gamma;\beta,\theta,\phi)\, ,
\label{mfromt}
\end{equation}

\noindent where $f_{{\cal Q}}$ is the quarkonium decay constant;
the $C^{j_1,\ j_2,\ J}_{m_1,m_2,M}$ are the well-known
Clebsch--Gordan coefficients; $\theta$, $\phi$ are the polar
and azimuthal angles specifying the direction of the
relative momentum between the two gluons, $2\bm{k}$,
in the glueball rest frame; $\beta$ is the modulus of the
relative momentum in units of the glueball mass,
$\beta=2|\bm{k}|/M_G$. Without giving more details on the
calculation of the hard scattering amplitudes $T$, let us only recall
that, at leading order in the strong coupling constant power expansion,
there are six Feynman graphs contributing to the process,
corresponding to all the possible ways the photon and the two gluons
can attach to the fermionic line.
The expressions of the helicity amplitudes
$M_{\lambda_\gamma,\lambda_1,\lambda_2;\lambda_{{\cal Q}}}$
are collected in Appendix \ref{appa}.

Following our approach \cite{Houra,Kessl}, the hadronic amplitude,
$A^{L,\ S,\ J}_{\lambda_\gamma,\lambda_G;\lambda_{{\cal Q}}}
(\theta_\gamma,\phi_\gamma)$, for the production
of a two-gluon bound state, $G(gg)$, with spin, orbital, total angular
momentum, and helicity $S$, $L$, $J$, $\lambda_G$ respectively,
is given by

\widetext

\begin{equation}
A^{L,\ S,\ J}_{\lambda_\gamma,\lambda_G;\lambda_{{\cal Q}}}
(\theta_\gamma,\phi_\gamma)=
2g_L^*\,\int^{1}_{-1}dz\ \frac{\Phi^*(z)}{\sqrt{1-z^2}}\
\lim_{\beta\to0}\frac{1}{\beta^L}\,\int\frac{d(\cos\theta)d\phi}
{4\pi}\ \sum_{\lambda_1,\lambda_2}\zeta^{L, S, J, \lambda_G}_{\lambda_1,
\lambda_2}(\theta,\phi)
M_{\lambda_\gamma,\lambda_1,\lambda_2;\lambda_{{\cal Q}}}
(\theta_\gamma,\phi_\gamma;\beta,\theta,\phi)\, ,
\label{afrom}
\end{equation}

\noindent where $g_L$ is the glueball decay constant,

\begin{equation}
g_L^* = \frac{1}{2}\left(\frac{2L+1}{2\pi M_G}\right)^{1/2}
\left(-\frac{2i}{M_G}\right)^L\ \frac{(2L+1)!!}{L!}
\left[\frac{d^L}{dr^L}R_L(r)\right]_{r=0}\, ;
\label{gl}
\end{equation}

\noindent $R_L(r)$ is the glueball radial wave function in momentum space;
$\Phi(z)$ is the glueball distribution amplitude,
which describes how the glueball longitudinal momentum (in the so-called
infinite momentum frame) is shared between its two constituent gluons
($x_{1,2} = (1\pm z)/2$ is the fractional longitudinal momentum carried
by gluon $1,\,2$ respectively), and 

\begin{equation}
\zeta^{L, S, J, \lambda_G}_{\lambda_1,\lambda_2}(\theta,\phi)=
C^{L\ S\ J}_{0\ \bar{\lambda}\ \bar{\lambda}}
C^{\ 1\ \ 1\ \ S}_{\lambda_1,-\lambda_2\ \bar{\lambda}}\,
e^{-i\lambda_G\phi}\,d^J_{\lambda_G,\bar{\lambda}}(\theta)\, ,
\label{zeta}
\end{equation}

\noindent where $\bar{\lambda}=\lambda_1-\lambda_2$.

It can be easily checked that Eq.~(\ref{afrom}) reduces to the usual
result of perturbative QCD for $L=0$ bound states, apart from some
trivial redefinitions of the decay constants.
In fact the same expression, with obvious appropriate modifications,
can be used for a $Q\bar{Q}$ bound state, and Eq.~(\ref{mfromt})
could be retrieved from Eq.~(\ref{afrom}) by substituting the gluons
with the $Q\bar{Q}$ pair, putting $L=0$, $S=J=1$, and using for the $J/\psi$
distribution amplitude the nonrelativistic limit,  $\Phi^{NR}(z)=
\delta(z)$.

Let us now consider in more detail the total glueball wave function,
$\Psi_G$. It must be completely symmetric under exchange of
the two gluons, $g_1 \longleftrightarrow g_2$.
$\Psi_G$ is factorized into the product of the color, spin, orbital
wave functions, and of the distribution amplitude,
$\Psi_G=\Phi_c\Phi_S\Phi_L\Phi(z)$.
Since the two gluons must be in a color singlet state, $\Phi_c$ is
symmetric. The combination of two $s=1$ particles can give
$S=0,1,2$, and $\Phi_S(g_1 \longleftrightarrow g_2)=(-1)^S \Phi_S$;
the orbital angular momentum component has symmetry $(-1)^L$;
finally, in our approach the distribution amplitude $\Phi(z)$ is
always taken symmetric in $z$. As a consequence, to have a
completely symmetric wave function, only states with even $L+S$
are allowed. We have explicitly checked that all amplitudes
with odd $L+S$ would require a nonvanishing,
antisymmetric distribution amplitude. We do not consider
this ``exotic'' possibility in the following. 
The parity of the glueball state is given by $P=(-1)^L$,
while its charge conjugation is $C=(-1)^{L+S}=+1$ for
all allowed states.

A (pure two-gluon) glueball bound-state $|J^{PC}\rangle$ is given
in the most general case by a linear combination of states with different
values of $L$, $S$ and same $J,P,C$ quantum numbers (only states with
even $L+S$ and $(-1)^L=P$ may contribute).
In Table \ref{jls} we show, for each glueball state with quantum numbers
$J^{PC}$ (up to $J=4$) all the $|L,S\rangle$ states with
$L\leq 4$ contributing.
We can put this in a concise form:

\begin{equation}
f_{J^{PC}}\,|J^{PC}\rangle =
 \sum_{S=0,1,2}\sum_{L=|J-S|}^{J+S}\frac{1}{4}
 \Bigl[1+(-1)^{L+S}\Bigr]\Bigl[1+P(-1)^L\Bigr]
 A_{L,S}\,f_L\,|L,S\rangle\,,
\label{jpcmix}
\end{equation}

\noindent where $f_{J^{PC}}$ is the decay constant of the physical state
and $A_{L,S}$ are (complex) mixing coefficients; whenever only one
$|L,S\rangle$ state contributes, we take $A_{L,S}\equiv 1$ and
$f_{J^{PC}}\equiv f_L$.
The expressions of the helicity amplitudes for all
$|L,S\rangle$ states of interest are reported in Appendix \ref{appb}.

Let us remark that in our approach the gluons
in the two-gluon bound state have an effective constituent mass,
$m_{1,2}=(M_G/2)\sqrt{(1\pm z)^2-\beta^2}$. This fact has two important
consequences: {\it i)} The gluons have three possible helicity states,
$\lambda_{1,2}=\pm1,0$. It is easy to check (see Appendix \ref{appb})
that all amplitudes with $\lambda_1=0$ and/or
$\lambda_2=0$ correctly vanish in the limit
$m_{1,2}\to 0$. {\it ii)} Yang's theorem states
that two massless spin 1 particles cannot bind to form a $J=1$ state,
or a state with odd $J$ and negative parity. In principle
this theorem does not apply to our model with constituent gluons;
in fact we find nonvanishing amplitudes for the $1^{++}$,
$1^{-+}$, $3^{-+}$ bound states;
however, all amplitudes involved correctly vanish in the
limit $m_{1,2}\to 0$.
The observation of two-gluon, $1^{++}$, $1^{-+}$, $3^{-+}$
bound states in radiative decays of quarkonium should 
indeed be an interesting test in favor of approaches, like ours,
involving massive constituent gluons \cite{corn}.

\section{Decay rates and angular distributions}
\label{results}

In the previous section we have presented all general results regarding the
helicity scattering amplitudes required for the calculation of
decay rates, photon angular distributions, etc.
In this section we first discuss the choice of the (nonperturbative)
glueball distribution amplitude, $\Phi(z)$. We will consider two
physically plausible and justified possibilities that, 
on one hand, give indications on the dependence of numerical
results on $\Phi(z)$ and that, on the other hand, are
such that the integrals in the expressions of the helicity amplitudes,
see Appendix~\ref{appb}, can be performed analytically.
After this discussion, we will present the expressions of the decay rate
and of the photon angular distribution for the process considered,
in terms of the helicity amplitudes.

Looking at the results of Appendix~\ref{appb}, one sees that all the
integrals appearing in the helicity amplitudes are of the following form:

\begin{equation}
\Psi_{l,m,n}(y) = 2\int^1_0\,dz\,\frac{\Phi(z)}{\sqrt{1-z^2}}\,
\frac{z^{2m}}{(1-z^2)^l(1-y^2z^2)^n}\, ,
\label{psint}
\end{equation}

\noindent
where $y=M_G^2/M_{{\cal Q}}^2$.
To perform numerical calculations, one needs an explicit expression for the
glueball distribution amplitude, $\Phi(z)$. This
function is largely unknown, given its nonperturbative origin.
The simplest possible option would be that of considering
the nonrelativistic limit, $\Phi^{NR}(z) = \delta(z)$.
This gives very simple analytical results for the helicity amplitudes;
however, it is generally considered inaccurate,
in particular for low mass bound states; therefore, we will not
consider furthermore this option in the rest of the paper.
Let us only notice that the reader can easily derive all useful results
for the nonrelativistic case by himself from Appendix~\ref{appb}.
We will then consider two more realistic options, based on results
obtained in the past years for the production/decay of light mesons
at high energies and largely adopted in the literature in various
forms: \\
 i) A generalized version of the so-called asymptotic
distribution amplitude \cite{Murgia},

\begin{equation}
 \Phi^{GAS}_{L}(z) = N_L(1-z^2)^{L+1}=\frac{\Gamma(L+5/2)}{\sqrt{\pi}
 \Gamma(L+2)}(1-z^2)^{L+1}\, ,
\label{gasy}
\end{equation}
 
\noindent where $\Gamma(z)$ is the well-known Euler Gamma function. \\
 ii) Following ref.s\cite{dzie,ji,huan}, a version of the asymptotic
distribution amplitude which suppresses end-point contributions,

\begin{eqnarray}
 \Phi^{AS}(z) & = & N_u\,(1-z^2)\,\exp\left[\,-u/(1-z^2)\,\right]
\nonumber \\
& = &
\frac{3}{2}\, \exp\left(\,u/2\,\right)\; \left\{\,u^2\,
K_{0}(u/2) + u (1-u)\, K_{1}(u/2)\,\right\}^{-1}\,(1-z^2)\;
\exp\left[\,-u/(1-z^2)\,\right]\, , 
\label{masy} 
\end{eqnarray}

\noindent where $u=M_G^2/(2b^2)$, $b$ is a hadronic scale parameter
and $K_{0}$ and $K_{1}$ are the well-known modified Bessel
functions of the second kind. Notice that when
$b\to 0$ (or better when $M_G \gg b$)
$\Phi^{AS}(z)\to \delta(z)$;
for light mesons, like e.g. the pion, and reasonable values of
$b\sim 0.5-1.0$ GeV, $\Phi^{AS}(z)$ is very similar to the well
known asymptotic distribution amplitude \cite{Brod}. 
 \\
With these choices for the glueball distribution amplitude,
all integrals $\Psi_{l,m,n}(y)$, Eq.~(\ref{psint}), can be
performed analytically. A more detailed discussion of these integrals
and a collection of useful results are presented in Appendix C.

The decay rate and the photon angular distribution for the process
${\cal Q}\to\gamma\, G(J^{PC})$ can be easily obtained
from the differential decay rate

\begin{equation}
 \frac{d\Gamma}{d\phi_\gamma\ d(\cos\theta_\gamma)} =
 \frac{1}{64\pi^2M_{{\cal Q}}}(1-y)\sum_{\lambda_{{\cal Q}},
 \lambda'_{{\cal Q}}}\sum_{\lambda_\gamma,\lambda_G}
 \rho_{\lambda_{{\cal Q}},\lambda'_{{\cal Q}}}({\cal Q})
 A_{\lambda_\gamma,\lambda_G;\lambda_{{\cal Q}}}(\theta_\gamma,
 \phi_\gamma)A^*_{\lambda_\gamma,\lambda_G;\lambda'_{{\cal Q}}}
 (\theta_\gamma,\phi_\gamma)\, ,
\label{dgam}
\end{equation}

\noindent where $\rho({\cal Q})$ is the helicity density matrix
of the quarkonium state, normalized so that ${\rm Tr}\rho({\cal Q})=1$.
$\rho({\cal Q})$ depends on the quarkonium production dynamics.
For typical production in high energy $\ell^+\ell^-$ colliders
all off-diagonal elements are vanishing,
$\rho_{\pm 1,\pm 1}({\cal Q})\simeq 1/2$,
$\rho_{00}({\cal Q})\simeq 0$; in the following applications we
will consider this case.

By using Eq.s~(\ref{ared}), (\ref{ared0}) one can easily see that
the total decay width can be written in the form

\begin{equation}
 \Gamma({\cal Q}\to\gamma\, G)=
 \frac{1}{16\pi M_{{\cal Q}}}(1-y)\,\frac{1}{3}\sum_{\lambda_\gamma,
 \lambda_G}|\hat{A}_{\lambda_\gamma,\lambda_G}|^2\, ,
\label{dgamtot}
\end{equation}

\noindent while the decay polar angular distribution, normalized to
unity, can be written as

\begin{equation}
 \frac{1}{\Gamma}\frac{d\Gamma}{d(\cos\theta_\gamma)}=
 \frac{3}{2}\frac{1}{\sum_{\lambda_\gamma,\lambda_G}
 |\hat{A}_{\lambda_\gamma,\lambda_G}|^2}
 \sum_{\lambda_{{\cal Q}},\lambda_\gamma,\lambda_G}
 \rho_{\lambda_{{\cal Q}},\lambda_{{\cal Q}}}({\cal Q})
 \Bigl[d^1_{\lambda_{{\cal Q}},\lambda_\gamma-\lambda_G}
 (\theta_\gamma)\Bigr]^2|\hat{A}_{\lambda_\gamma,\lambda_G}|^2\, .
\label{dgamuni}
\end{equation}

If $|\hat{A}_{+1,0}| \neq 0$ we can define, for a glueball state
with total angular momentum $J$,

\begin{equation}
r_i^J = \frac{|\hat{A}_{+1,+i}|^2}{|\hat{A}_{+1,0}|^2}\, ,
\label{rj}
\end{equation}

\noindent where $i=1,2$ and, since $|\lambda_G|\leq J$, $r_i^J= 0$
if $J<i$. Then, using Eq.~(\ref{dgamuni}), one finds

\begin{equation}
 \frac{1}{\Gamma^J}\frac{d\Gamma^J}{d(\cos\theta_\gamma)}=
 \frac{3}{4}\frac{1-2r_1^J+r_2^J}{1+r_1^J+r_2^J}
 \Biggl\{\,\frac{1+r_2^J}{1-2r_1^J+r_2^J}-\rho_{1,1}({\cal Q})
 -\Bigl[1-3\rho_{1,1}({\cal Q})\Bigr]\cos^2\theta_\gamma\,\Biggr\}\, .
\label{dgmarj0}
\end{equation}

If $|\hat{A}_{+1,0}|=0$ and $|\hat{A}_{+1,1}|\neq 0$ we may in turn
define

\begin{equation}
r_2^J = \frac{|\hat{A}_{+1,+2}|^2}{|\hat{A}_{+1,1}|^2}\, ,
\label{rj1}
\end{equation}

\noindent and Eq.~(\ref{dgmarj0}) simplifies to

\begin{equation}
 \frac{1}{\Gamma^J}\frac{d\Gamma^J}{d(\cos\theta_\gamma)}=
 \frac{3}{4}\frac{r_2^J-2}{1+r_2^J}
 \Biggl\{\,\frac{r_2^J}{r_2^J-2}-\rho_{1,1}({\cal Q})
 -\Bigl[1-3\rho_{1,1}({\cal Q})\Bigr]\cos^2\theta_\gamma\,\Biggr\}\, .
\label{dgmarj1}
\end{equation}

\section{Numerical results}
\label{numris}

Since our approach cannot predict the glueball decay constants, $f_L$,
we are not able to give reliable estimates for absolute quantities
like decay rates, unless using information from other nonperturbative
approaches. Another source of uncertainty is that, as we have seen,
even considering only pure two-gluon states, some of the bound states
with given quantum numbers, $J^{PC}$, may result from the mixing of
several $L,S$ states, with unknown mixing coefficients.
Although for weakly bound systems it is usually assumed that amplitudes
corresponding to larger values of $L$ are suppressed,
the overall contribution of each $L,S$ state may be modified by the
inclusion of the mixing coefficients.
Therefore, in this paper we limit ourself to give predictions
separately for each $L,S$ contribution and for quantities
that do not depend on the glueball decay constants, i.e. the
angular distribution of the decay products and the branching ratios
$R=\Gamma(\Upsilon\to \gamma\, G)/\Gamma(J/\psi\to \gamma\, G)$.
For this last quantity some experimental information is available
concerning possible glueball candidates.
Both these observables can in principle give useful information
regarding the glueball distribution amplitude and, for a given $J^{PC}$
state, regarding the relative weight of the various $L,S$
amplitudes contributing. 

\subsection{Predictions for the ratio $\bm{R=\Gamma(\Upsilon \to
 \gamma\, G)/\Gamma(J/\psi \to \gamma\, G)}$} \label{numR}

Let us consider the ratio $R=\Gamma(\Upsilon \rightarrow \gamma\, G)/
\Gamma(J/\psi \rightarrow \gamma\,G)$. $R$ is independent of the
value of the glueball decay constant. On the other hand, it depends
on the quarkonium decay constants $f_\psi$, $f_\Upsilon$.
{}From Eq.~(\ref{dgamtot}) and the results of Appendix \ref{appb}
we easily see that

\begin{equation}
R=\frac{\Gamma(\Upsilon \rightarrow \gamma\, G)}
{\Gamma(J/\psi \rightarrow \gamma\,G)}=
\frac{1-y_{_\Upsilon}}{1-y_{_\psi}}\,
\left(\,\frac{\alpha_s(M_\Upsilon^2)}{\alpha_s(M_\psi^2)}\,\right)^2\,
\left(\frac{\,f_\Upsilon}{f_\psi}\,\right)^2\,
\left(\,\frac{M_\psi}{M_\Upsilon}\,\right)^3\,
\frac{F(\,y_{_\Upsilon})}{F(\,y_{_\psi})} \,,
\label{rfm}
\end{equation}

\noindent
where $y_{{\cal Q}}=(M_G/M_{{\cal Q}})^2$ and $F(\,y_{_{\Upsilon,\psi}})$
includes the rest of the contribution from the helicity amplitudes
not explicitly shown.
We can make use of the well known leading order result

\begin{equation}
\Gamma({\cal Q}(^{3}S_{1})\to e^{+}e^{-}) =
\frac{32}{27}\,\pi\alpha^{2}\,\frac{f^{\,2}_{{\cal Q}}}{M_{{\cal Q}}}\,,
\label{gee}
\end{equation}

\noindent
and of the leading order expression for the strong coupling constant

\begin{equation}
\alpha_s(Q^2) = \frac{12\pi}{(11n_c-2n_f)\ln(Q^2/\Lambda^2)}\,,
\label{alfas}
\end{equation}

\noindent
with $n_c=3$, $n_f=4,5$ for $Q=M_\psi,M_\Upsilon$ respectively, and
$\Lambda=0.2$ GeV. We then find

\begin{equation}
R=\frac{\Gamma(\Upsilon\to e^{+}e^{-})}
{\Gamma(J/\psi\to e^{+}e^{-})}\,
\left(\,\frac{M_\psi}{M_\Upsilon}\,\right)^{2}\,
\frac{1-y_{_\Upsilon}}{1-y_{_\psi}}\,
\left(\,\frac{25\ln(M_\psi^2/\Lambda^2)}
{23\ln(M_\Upsilon^2/\Lambda^2)}\,\right)^2\,
\frac{F(y_{_\Upsilon})}{F(y_{_\psi})}\,.
\label{rmgee}
\end{equation}

\noindent
We make also use of the experimental results \cite{pdg}

\begin{eqnarray}
\Gamma^{\text{exp}}(J/\psi \to e^{+}e^{-}) &=&
  5.26 \pm 0.37~\text{KeV} \, , \nonumber \\
\Gamma^{\text{exp}}(\Upsilon \to e^{+}e^{-}) &=&
  1.32 \pm 0.04 \pm 0.03~\text{KeV} \, .
\label{geexp}
\end{eqnarray}

In Table \ref{rtab} we present our results for the ratio $R$,
for several glueball bound states. For each state, the mass is chosen
according to the most favored experimental candidate available
\cite{exp,zou,pdg}.
Due to its interest, we also consider one of the states forbidden
by Yang's theorem, i.e. the $1^{++}$ state, for which 
we have indicatively taken $M_{1^{++}}=2.34$ GeV.

For each state, we have also separately considered all possible
$|L,S\,\rangle$ states contributing. This could be useful, together
with the study of the decay angular distributions, to get information
on the unknown mixing coefficients.
To study the dependence of our results on the choice of
the glueball distribution amplitude, we
present results for the generalized asymptotic distribution amplitude
$\Phi^{GAS}_{L}$, Eq.~(\ref{gasy}), and for the modified
asymptotic distribution amplitude $\Phi^{AS}$, Eq.~(\ref{masy}),
using in this case two indicative values for the parameter
$b$, $b=0.5$ GeV and $b=1.0$ GeV. 

For two of the most accredited glueball candidates \cite{pdg},
i.e. the resonances $f_{0}(1710)$ and $f_{J}(2220)$ ($J=2$ or 4),
some experimental results on branching ratios are available:

\begin{eqnarray}
B^{\text{exp}}(J/\psi \to \gamma\, f_{0}(1710))\times 
 B^{\text{exp}}(f_{0}(1710) \to K^{+}K^{-}) &=&
 8.5^{+1.2}_{-0.9}\times10^{-4} \, , \nonumber\\
B^{\text{exp}}(J/\psi \to \gamma\, f_{J}(2220))\times 
 B^{\text{exp}}(f_{J}(2220) \to K^{+}K^{-}) &=&
 (8.1 \pm 3.0)\times10^{-5} \, , \\
B^{\text{exp}}(\Upsilon \to \gamma\, f_{0}(1710))\times 
 B^{\text{exp}}(f_{0}(1710) \to K^{+}K^{-}) &<&
 2.6\times10^{-4} \, , \nonumber\\
B^{\text{exp}}(\Upsilon \to \gamma\, f_{J}(2220))\times 
 B^{\text{exp}}(f_{J}(2220) \to K^{+}K^{-}) &<& 
 1.5\times10^{-5} \, .
\label{f0fj}
\end{eqnarray}

Taking the lowest possible value for the $J/\psi$ case, this results
in upper bounds for the ratio $R$, which are reported in Table \ref{rtab}.

{}From Table \ref{rtab} we can see that, apart from two
cases (i.e., the $L=2$, $S=0$ contribution to the $2^{++}$ state
and the $L=4$, $S=0$ contribution to the $4^{++}$ state), the results
show relatively little dependence on the glueball distribution amplitude,
differing at most by a factor of two to three.
Our estimates are consistent with the experimental upper limits,
when available; for the $4^{++}$ state the results are very close to
(and in some case slightly larger than) the upper limit, while in all
other cases they are considerably smaller. For a given state, there are
remarkable differences between the possible $L,S$ contributions, which
could be useful once more experimental information on the ratio $R$
will hopefully be available. Complementary information can be found
by looking at the photon angular distributions.

Regarding the scalar, $0^{++}$ glueball state, since the $f_0(1500)$
is also reported to be a possible candidate as an alternative to the
$f_0(1710)$, we have also considered the case $M_{0^{++}}=1.5$
GeV. The value of the ratio $R$ correspondingly decreases (increases)
by 10-20\% (20-30\%) at most for the $L=0$ ($L=2$) contribution.

In spite of the lack of detailed experimental information, we have
investigated the dependence of the ratio $R$ on the glueball mass
for the other states as well, using other possible mass
attributions \cite{zou}.
While the $0^{-+}$, $1^{++}$, $2^{++}$ states show
on the overall a dependence smaller than,
or comparable to, that of the scalar case, the
$2^{-+}$, $3^{++}$, $4^{++}$ states show a stronger sensitivity.
As an example, a 10\% change in the mass of the $4^{++}$ state
induces a change of up to a factor 2-3 in the ratio $R$, the sign
and the value of the change depending on the particular $L$ contribution
and glueball wave function considered.
This sensitivity could be of some usefulness in discriminating among possible
candidates in the mass range where our model is applicable.

\subsection{Photon angular distributions}

\begin{center}
\begin{figure}[t]
\epsfig{file=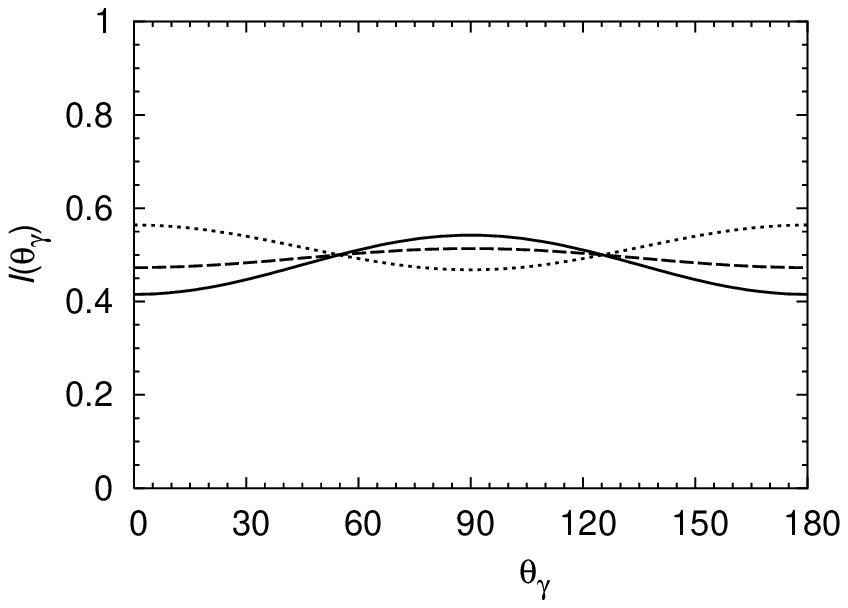,width=11.6truecm}
\caption{Normalized photon angular distribution, $I(\theta_\gamma)=
(1/\Gamma)\,d\,\Gamma/d(\cos\theta_\gamma)$, for the $\Upsilon\to\gamma
\,G(1^{++})$ decay and different choices of
the glueball distribution amplitude (DA), $\Phi(z)$:
asymptotic DA (ASY DA), $b=0.5$ GeV (solid);
ASY DA, $b=1.0$ GeV(dashed); generalized asymptotic (GAS) DA (dotted). }
\label{u1pp}
\end{figure}
\end{center}

\begin{center}
\begin{figure}[t]
\epsfig{file=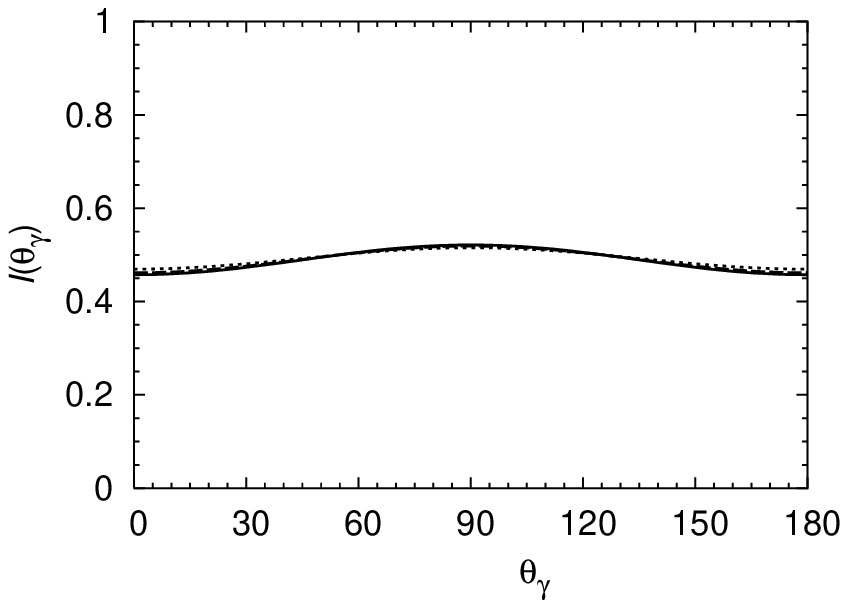,width=11.6truecm}
\caption{Normalized photon angular distribution, $I(\theta_\gamma)=
(1/\Gamma)\,d\,\Gamma/d(\cos\theta_\gamma)$, for the $J/\psi\to\gamma
\,G(1^{++})$ decay and different choices of
the glueball distribution amplitude (DA), $\Phi(z)$:
asymptotic DA (ASY DA), $b=0.5$ GeV (solid);
ASY DA, $b=1.0$ GeV(dashed); generalized asymptotic (GAS) DA (dotted). }
\label{j1pp}
\end{figure}
\end{center}

\begin{center}
\begin{figure}[t]
\epsfig{file=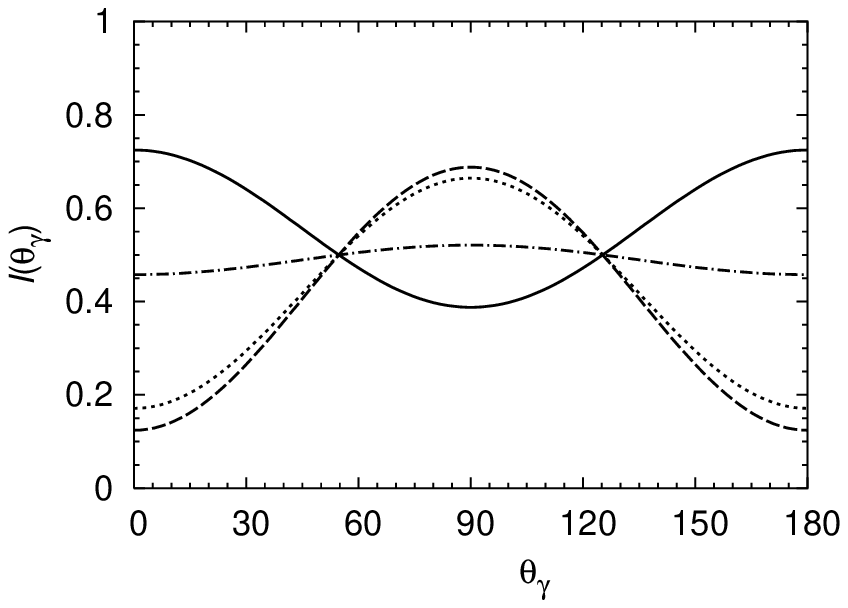,width=11.6truecm}
\caption{Normalized photon angular distribution, $I(\theta_\gamma)=
(1/\Gamma)\,d\,\Gamma/d(\cos\theta_\gamma)$, for the $\Upsilon\to\gamma
\,G(2^{-+})$ decay and different $|L,S\,\rangle$ glueball states:
$L=1$, $S=1$ (solid); $L=3$, $S=1$, ASY DA with $b=0.5$ GeV (dashed),
ASY DA with $b=1.0$ GeV (dotted), GAS DA (dot-dashed). Notice that for
the $L=1$, $S=1$ case $I(\theta_\gamma)$ is independent of the DA adopted.}
\label{u2mp}
\end{figure}
\end{center}

\begin{center}
\begin{figure}[t]
\epsfig{file=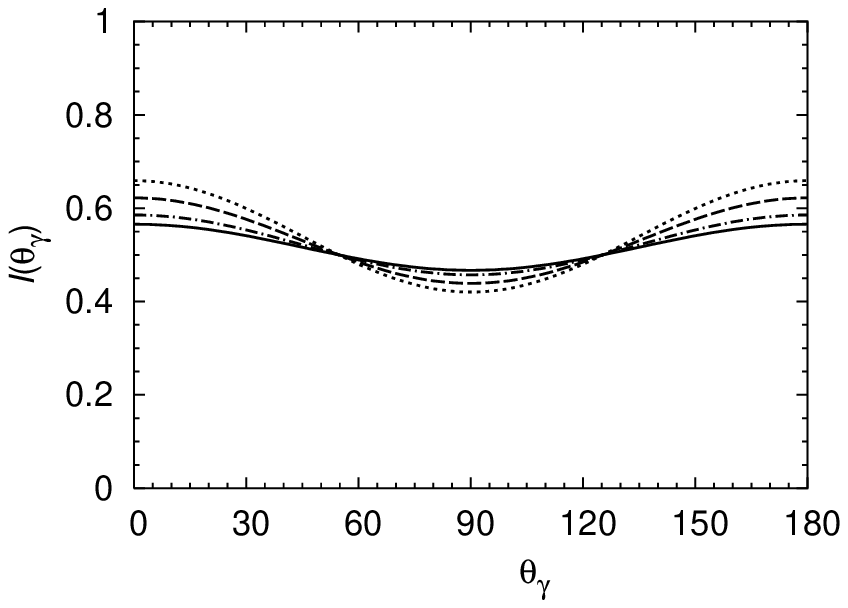,width=11.6truecm}
\caption{Normalized photon angular distribution, $I(\theta_\gamma)=
(1/\Gamma)\,d\,\Gamma/d(\cos\theta_\gamma)$, for the $J/\psi\to\gamma
\,G(2^{-+})$ decay and different $|L,S\,\rangle$ glueball states:
$L=1$, $S=1$ (solid); $L=3$, $S=1$, ASY DA with $b=0.5$ GeV (dashed),
ASY DA with $b=1.0$ GeV (dotted), GAS DA (dot-dashed). Notice that for
the $L=1$, $S=1$ case $I(\theta_\gamma)$ is independent of the DA adopted.}
\label{j2mp}
\end{figure}
\end{center}

\begin{center}
\begin{figure}[t]
\epsfig{file=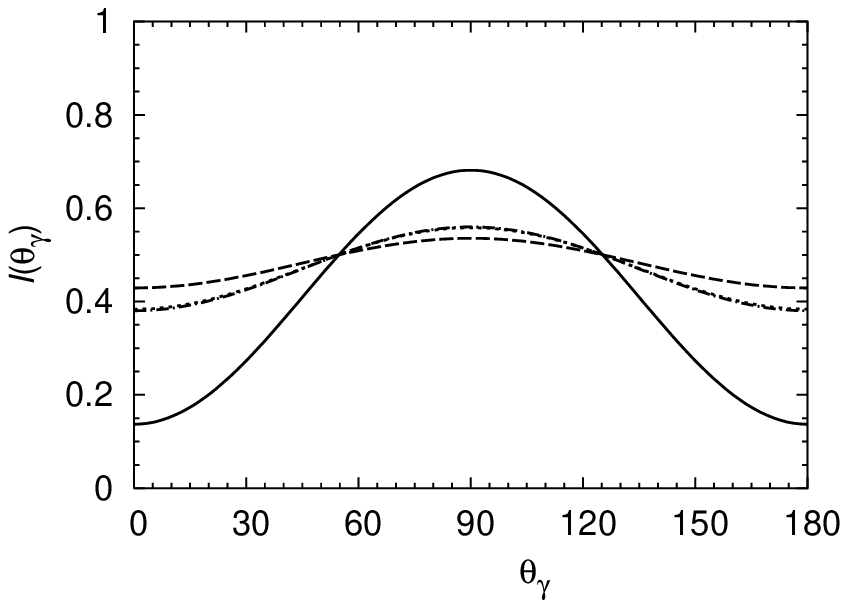,width=11.6truecm}
\caption{Normalized photon angular distribution, $I(\theta_\gamma)=
(1/\Gamma)\,d\,\Gamma/d(\cos\theta_\gamma)$, for the $\Upsilon\to\gamma
\,G(3^{++})$ decay and different $|L,S\,\rangle$ glueball states:
$L=2$, $S=2$ (solid); $L=4$, $S=2$, ASY DA with $b=0.5$ GeV (dashed),
ASY DA with $b=1.0$ GeV (dotted), GAS DA (dot-dashed). Notice that for
the $L=2$, $S=2$ case $I(\theta_\gamma)$ is independent of the DA adopted.}
\label{u3pp}
\end{figure}
\end{center}

\begin{center}
\begin{figure}[t]
\epsfig{file=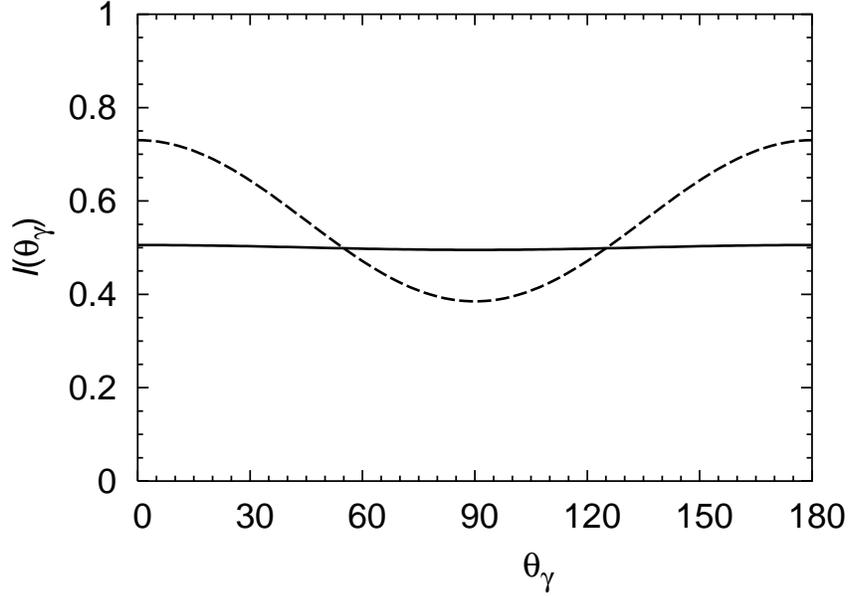,width=11.6truecm}
\caption{Normalized photon angular distribution, $I(\theta_\gamma)=
(1/\Gamma)\,d\,\Gamma/d(\cos\theta_\gamma)$, for the $J/\psi\to\gamma
\,G(3^{++})$ decay and different $|L,S\,\rangle$ glueball states,
using the ASY DA with $b=0.5$ GeV:
$L=2$, $S=2$ (solid); $L=4$, $S=2$ (dashed). Notice that for the
$L=2$, $S=2$ case $I(\theta_\gamma)$ is independent of the DA adopted,
while for the $L=4$, $S=2$ case the three choices here adopted give almost
indistinguishable results.}
\label{j3pp}
\end{figure}
\end{center}

\begin{center}
\begin{figure}[t]
\epsfig{file=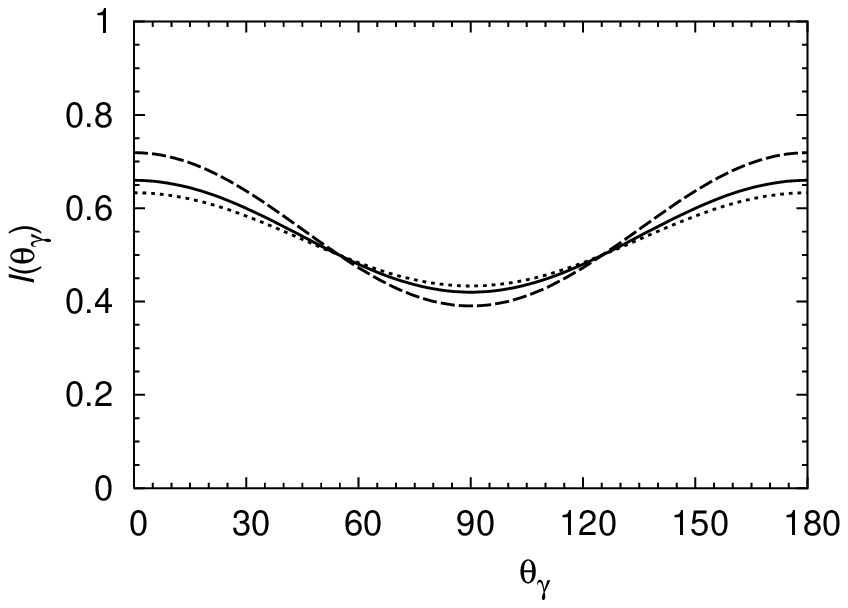,width=11.6truecm}
\caption{Normalized photon angular distribution, $I(\theta_\gamma)=
(1/\Gamma)\,d\,\Gamma/d(\cos\theta_\gamma)$, for the $\Upsilon\to\gamma
\,G(4^{++})$ decay and different $|L,S\,\rangle$ glueball states, using
the asymptotic DA (ASY DA), with $b=0.5$ GeV:
$L=2$, $S=2$ (solid); $L=4$, $S=0$ (dashed); $L=4$, $S=2$ (dotted).
Use of the ASY DA with $b=1.0$ GeV, or of the GAS DA,
leads to very similar results.}
\label{u4ppa}
\end{figure}
\end{center}

\begin{center}
\begin{figure}[t]
\epsfig{file=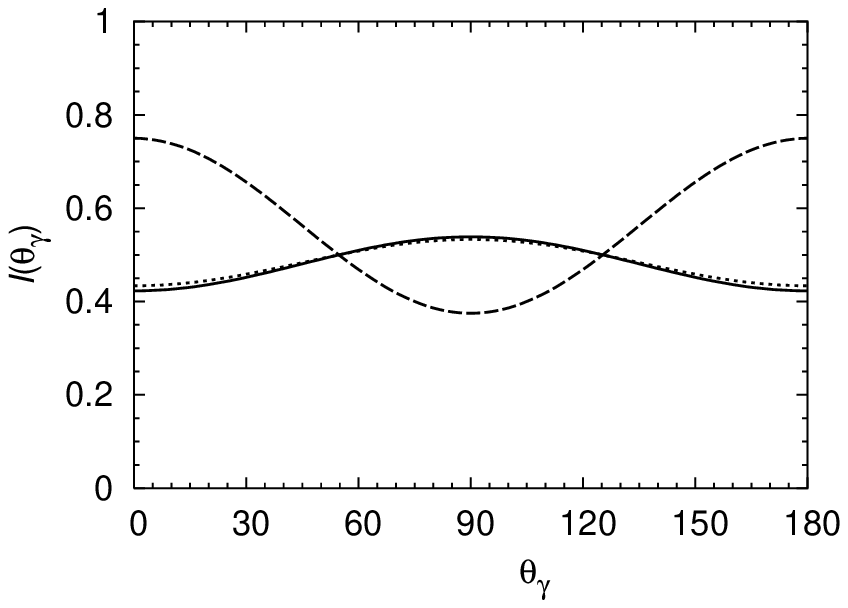,width=11.6truecm}
\caption{Normalized photon angular distribution, $I(\theta_\gamma)=
(1/\Gamma)\,d\,\Gamma/d(\cos\theta_\gamma)$, for the $J/\psi\to\gamma
\,G(4^{++})$ decay and different $|L,S\,\rangle$ glueball states, using
the asymptotic DA (ASY DA), with $b=0.5$ GeV:
$L=2$, $S=2$ (solid); $L=4$, $S=0$ (dashed); $L=4$, $S=2$ (dotted).
Use of the ASY DA with $b=1.0$ GeV, or of the GAS DA,
leads to very similar results.}
\label{j4ppa}
\end{figure}
\end{center}

Using the results of Eq.s~(\ref{dgmarj0}), (\ref{dgmarj1}), we present
in Fig.s~\ref{u1pp}-\ref{j4ppa} the normalized photon angular distribution,
$I(\theta_\gamma)=(1/\Gamma)\,d\,\Gamma/d(\cos\theta_\gamma)$,
for the decays $\Upsilon,\,J/\psi\to\gamma\,G$, considering the
glueball candidates of Table \ref{rtab}.
For each $J^{PC}$ state we give the angular distribution for all
possible $L,S$ contributions separately, and for the three different
choices of the glueball distribution amplitude.
Notice that for the $0^{++}$ and $0^{-+}$ states, where for each $L,S$
contribution there is only one nonvanishing amplitude
(the one with $\lambda_G=0$),
$I(\theta_\gamma)=(3/8)(1+\cos^2\theta_\gamma)$, which
gives no useful information.
There are also some cases in which, for a pure $L,S$ contribution,
$I(\theta_\gamma)$ is independent of the glueball distribution amplitude.
This happens when, at fixed $L,S$ values, all nonvanishing amplitudes
with different values of $\lambda_G$ have the same functional form of
the integrals over the variable $z$. More precisely, this situation applies
to the following cases: the $L=0$, $S=2$ contribution to the $2^{++}$
state; the $L=1$, $S=1$ contribution to the $2^{-+}$ state;
the $L=2$, $S=2$ contribution to the $3^{++}$ state; the
$L=2$, $S=2$ contribution to the $4^{++}$ state; it also applies to
the $4^{-+}$ state, not considered in Table \ref{rtab}.

Regarding Fig.s~\ref{u1pp}-\ref{j4ppa} one can make the following remarks:

\noindent
1) Apart from the cases where, for the reasons discussed above,
there is no dependence on the glueball distribution amplitude
at all, practically all remaining cases show little or
negligible dependence on $\Phi(z)$. The only remarkable case is the
$L=3$, $S=1$ contribution to $\Upsilon\to\gamma\,G(2^{-+})$,
where there is a sizable difference between the results obtained
with the generalized asymptotic DA and those with the modified
asymptotic DA.

\noindent
2) Regarding the possibility of discriminating among the different
$|L,S\rangle$ contributions to a given $J^{PC}$ state, the overall
situation looks more interesting. There are remarkable differences
in the $2^{++}$ case, for the $J/\psi$ decay; in the $2^{-+}$ case,
for the $\Upsilon$ decay; in the $3^{++}$ case, for both $\Upsilon$ and
$J/\psi$ decays; in the $4^{++}$ case, for the $J/\psi$ decay.

\noindent
3) The $2^{-+}$, $3^{++}$, $4^{++}$ cases are those that, on the overall,
show the most interesting effects when comparing the $\Upsilon$ and
$J/\psi$ decays.

\section{Conclusions}
\label{concl}

In this paper we have presented a detailed derivation of the
helicity amplitudes, decay widths, and angular distributions
for two-gluon glueball production in radiative quarkonium
decays. To this end, we have further extended an approach
previously developed and applied by some of us to different
high-energy hadronic processes. Given the limited knowledge of
some of the nonperturbative ingredients entering the calculation
(the glueball distribution amplitude and decay constant; the
mixing coefficients among different $L,S$ contributions) we
limited ourself to presenting some indicative numerical results
for the decay widths and angular distributions in the
production of possible glueball candidates in a pure $L,S$ configuration,
adopting two different  models for the
glueball distribution amplitude. These results are mainly intended
to illustrate the potentiality of our approach in discriminating
among different glueball candidates and their masses,
quantum number attributions, etc., once it is complemented by experimental
information and other theoretical, nonperturbative inputs.
We have tried to present in detail all analytical results,
so that they could be used and even further extended by the
interested reader.

An interesting feature of our model is that it implies the use of
massive constituent gluons, so that two-gluon bound states
can escape Yang's theorem and form $1^{++}$, $1^{-+}$, $3^{-+}$
states, whose observation in the mass range considered here would
then be a strong argument in favor of approaches, like ours,
involving constituent gluons.

We conclude by stressing that our approach can be
generalized to include light $q\bar{q}$ pair (mesonic) production
and possibly exotic states made of a diquark-antidiquark pair.
This could be a necessary step if one would consider more realistic
situations, where the observed resonances may result
from the mixing of $q\bar{q}$, $gg$, etc. components.
Let us also remark that a similar approach could be extended
(and in fact this has been already done, at least in part)
to the case of hadronic resonance production in photon-photon
collisions. The completion of this unified approach would certainly
be of great interest for a deeper understanding of hadronic
structure and spectroscopy in the mass range of a few GeV/$c^2$.

\acknowledgments

We are grateful to P. Kessler for a critical reading of the manuscript
and to G.M. Pruna for some help and useful comments regarding Appendix C.
J.P. thanks INFN, Sezione di Cagliari, for the very warm hospitality
shown to him.
M.M. and F.M. are partially supported by M.I.U.R. (Ministero
dell'Istruzione, dell'Universit\`a e della Ricerca)
under Cofinanziamento P.R.I.N. 2001.

\appendix

\section{Helicity amplitudes for the process
$\bm{J/\psi,\Upsilon\to\gamma\, g^*g^*}$}
\label{appa}

\begin{center}
\begin{figure}[t]
\epsfig{figure=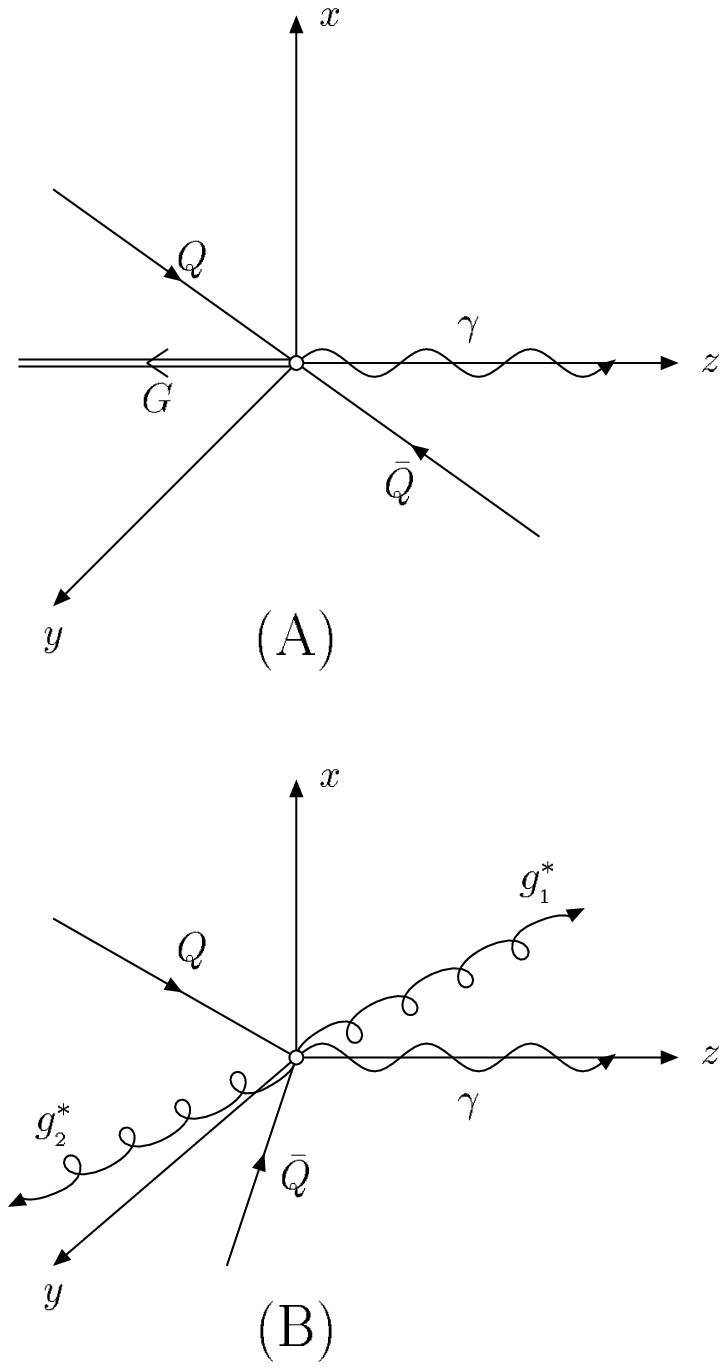,width=9truecm}
\vspace*{8pt}
\caption{Kinematical configuration for the process $Q\bar{Q}\to\gamma\,G$ 
in the quarkonium rest frame (A), and for the elementary partonic process
$Q\bar{Q}\to\gamma\,g^*\,g^*$ in the glueball rest frame (B).}
\label{kin}
\end{figure}
\end{center}

In this Appendix we present the expressions of the helicity amplitudes,
$M_{\lambda_\gamma,\lambda_1,\lambda_2;\lambda_{{\cal Q}}}$,
for the process ${\cal Q}\to\gamma\, g^*g^*$, with the kinematics defined in
the rest frame of ${\cal Q}$, see Fig.~\ref{kin}(A).
Notice that according to Eq.s~(\ref{ared}),~(\ref{ared0}), it is
sufficient to evaluate the helicity amplitudes in the simple
kinematical configuration $\theta_\gamma=\phi_\gamma=0$.
Let us recall that, while the amplitudes
are given in the quarkonium rest frame, the angles $\theta$
and $\phi$ specify the direction of the two-gluon relative momentum
in the glueball rest frame, see Fig.~\ref{kin}(B).
A simple Lorentz boost along the $z$-axis connects the two reference
frames. We further notice that, due to parity invariance,
only 27 out of all 54 amplitudes are in principle independent.
Therefore in the following relations we fix $\lambda_\gamma=+1$.

Let us define

\begin{equation}
K=\frac{128\,\pi^{3/2}\sqrt{\alpha}\alpha_s f_{{\cal Q}}}{9M_{{\cal Q}}}
 \frac{1}{[1+y(\beta^2-z^2)]^2-(1-y)^2(\beta\cos\theta-z)^2}\, ,
\label{km}
\end{equation}

\noindent where $y=M_G^2/M_{{\cal Q}}^2$.
We have for the helicity amplitudes $M_{\lambda_\gamma,
\lambda_1,\lambda_2;\lambda_{{\cal Q}}}(\theta,\phi)$ :

\begin{equation}
M_{+,\pm,\pm;+}=-K\Bigl\{(1-\beta^2-z^2)(1+\cos^2\theta)\mp
 2y\beta[(1\mp\beta)^2-z^2]+2z\{2\beta\pm y[(1\mp\beta)^2-z^2]\}
 \cos\theta\Bigr\}\,,
\end{equation}

\begin{equation}
M_{+,\pm,\mp;+}=-K\sin^2\theta(1+\beta^2-z^2)\,,
\end{equation}

\begin{equation}
M_{+,\pm,0;+}=\mp\sqrt{2}\,K\sqrt{(1-z)^2-\beta^2}\,\sin\theta
 \Bigl[(\cos\theta\pm yz)(1+z)-(1-y)\beta\mp y\beta^2\Bigr]\,,
\end{equation}

\begin{equation}
M_{+,0,\pm;+}=\pm\sqrt{2}\,K\sqrt{(1+z)^2-\beta^2}\,\sin\theta
 \Bigl[(\cos\theta\pm yz)(1-z)+(1-y)\beta\pm y\beta^2\Bigr]\,,
\end{equation}

\begin{equation}
M_{+,0,0;+}=2\,K\sqrt{(1+z)^2-\beta^2}\sqrt{(1-z)^2-\beta^2}\,\sin^2\theta\,,
\end{equation}

\begin{equation}
M_{+,\pm,\pm;0}=-\sqrt{2}\,K\,y^{1/2}\,e^{-i\phi}\sin\theta
\Bigl[(1-\beta^2-z^2)\cos\theta+z\{2\beta\pm y[(1\mp\beta)^2-z^2]\}\Bigr\}\,,
\end{equation}

\begin{equation}
M_{+,\pm,\mp;0}=\mp\sqrt{2}\,K\,y^{1/2}\,e^{-i\phi}\sin\theta
 (1\mp\cos\theta)(1+\beta^2-z^2)\,,
\end{equation}

\begin{equation}
M_{+,\pm,0;0}=\mp K\,y^{1/2}\sqrt{(1-z)^2-\beta^2}\,e^{-i\phi}
 (1\mp\cos\theta)\Bigl[(1+z)(1+yz\pm 2\cos\theta)
 \mp(1-y)\beta-y\beta^2\Bigr]\,,
\end{equation}

\begin{equation}
M_{+,0,\pm;0}=\pm K\,y^{1/2}\sqrt{(1+z)^2-\beta^2}\,e^{-i\phi}
 (1\pm\cos\theta)\Bigl[(1-z)(1-yz\mp 2\cos\theta)
 \mp(1-y)\beta-y\beta^2\Bigr]\,,
\end{equation}

\begin{equation}
M_{+,0,0;0}=-2\sqrt{2}\,K\,y^{1/2}\sqrt{(1+z)^2-\beta^2}
 \sqrt{(1-z)^2-\beta^2}\,e^{-i\phi}\sin\theta\cos\theta\,,
\end{equation}

\begin{equation}
M_{+,\pm,\pm;-}=-Ky\,e^{-i2\phi}\sin^2\theta(1-\beta^2-z^2)\,,
\end{equation}

\begin{equation}
M_{+,\pm,\mp;-}=-Ky\,e^{-i2\phi}(1\mp\cos\theta)^2(1+\beta^2-z^2)\,,
\end{equation}

\begin{equation}
M_{+,\pm,0;-}=-\sqrt{2}\,Ky\sqrt{(1-z)^2-\beta^2}\,e^{-i2\phi}
 (1+z)\sin\theta(1\mp\cos\theta)\,,
\end{equation}

\begin{equation}
M_{+,0,\pm;-}=-\sqrt{2}\,Ky\sqrt{(1+z)^2-\beta^2}\,e^{-i2\phi}
 (1-z)\sin\theta(1\pm\cos\theta)\,,
\end{equation}

\begin{equation}
M_{+,0,0;-}=-2Ky\sqrt{(1+z)^2-\beta^2}\sqrt{(1-z)^2-\beta^2}\,
 e^{-i2\phi}\sin^2\theta\,.
\end{equation}

\section{Helicity amplitudes for the process
$\bm{{\cal Q}\to\gamma\, G(L,S,J)}$}
\label{appb}

Using Eq.s~(\ref{afrom}),(\ref{gl}),(\ref{zeta}),
 and the results of Appendix \ref{appa},
we here present the expressions of the helicity amplitudes
$A^{L,\;S,\;J}_{\lambda_\gamma,\lambda_G;\lambda_{{\cal Q}}}
(\theta_\gamma=\phi_\gamma=0)$, for all glueball states up to
$J=4$ and $L\leq 4$.
The full angular dependence of the
amplitudes can be obtained from Eq.~(\ref{ared}). Moreover, due to parity
invariance, $A^{L,\;S,\;J}_{-\lambda_\gamma,-\lambda_G;-\lambda_{{\cal Q}}}
=(-1)^{L-S+\lambda_{{\cal Q}}-(\lambda_\gamma-\lambda_G)}\,
A^{L,\;S,\;J}_{\lambda_\gamma,\lambda_G;\lambda_{{\cal Q}}}$.
We here present only independent amplitudes, by fixing $\lambda_\gamma=+1$.
For each $|J^{PC}\rangle$ state we consider all possible $|L,S\rangle$
amplitudes contributing, according to Table \ref{jls}.

\vspace{12pt}

\noindent $\bm{J^{PC}=0^{++}}$

\begin{equation}
A^{0,\;0,\;0}_{1,\;0;\;1}=
 -\frac{256\sqrt{2}}{9\sqrt{3}}\pi\sqrt{\alpha}\alpha_s
 \frac{f_{{\cal Q}}}{M_{{\cal Q}}}\frac{|R_0(0)|}{M_G^{1/2}}
 \int_{-1}^{1}\!dz\,\frac{\Phi_N(z)}{\sqrt{1-z^2}}\frac{1}{1-y^2z^2}\,,
\end{equation}

\begin{eqnarray}
A^{2,\;2,\;0}_{1,\;0;\;1}&=&\frac{1024}{9\sqrt{3}}\pi\sqrt{\alpha}\alpha_s
 \frac{f_{{\cal Q}}}{M_{{\cal Q}}}\frac{|R_2''(0)|}{M_G^{5/2}}
 \int_{-1}^{1}\!dz\,\frac{\Phi_N(z)}{\sqrt{1-z^2}}
 \frac{1}{(1-z^2)^2(1-y^2z^2)^3}\nonumber\\
 &\times&\!\!\!\Bigl\{\,1+13y+y^2+(3+y-8y^2-34y^3+3y^4)z^2\nonumber\\
 &&-\,y^2(9+2y-31y^2-5y^3)z^4+5y^2(2-3y)z^6\,\Bigr\}\,.
\end{eqnarray}

\vspace{12pt}

\noindent $\bm{J^{PC}=0^{-+}}$

\begin{equation}
A^{1,\;1,\;0}_{1,\;0;\;1}=i\frac{512}{9}\pi\sqrt{\alpha}\alpha_s
 \frac{f_{{\cal Q}}}{M_{{\cal Q}}}\frac{|R_1'(0)|}{M_G^{3/2}}\,
 y\int_{-1}^{1}\!dz\,\frac{\Phi_N(z)}{\sqrt{1-z^2}}\,
 \frac{3-(1+4y+y^2)z^2+3y^2z^4}{(1-z^2)(1-y^2z^2)^2}\,.
\end{equation}

\vspace{12pt}

\noindent $\bm{J^{PC}=1^{++}}$

\begin{equation}
A^{2,\;2,\;1}_{1,\;0;\;1}=
 -\frac{2560\sqrt{2}}{3\sqrt{3}}\pi\sqrt{\alpha}\alpha_s
 \frac{f_{{\cal Q}}}{M_{{\cal Q}}}\frac{|R_2''(0)|}{M_G^{5/2}}\,y
 \int_{-1}^{1}\!dz\,\frac{\Phi_N(z)}{\sqrt{1-z^2}}\,
 \frac{1}{(1-z^2)^2(1-y^2z^2)}\,,
\end{equation}

\begin{eqnarray}
A^{2,\;2,\;1}_{1,\;1;\;0}&=&
 -\frac{512\sqrt{2}}{3\sqrt{3}}\pi\sqrt{\alpha}\alpha_s
 \frac{f_{{\cal Q}}}{M_{{\cal Q}}}\frac{|R_2''(0)|}{M_G^{5/2}}\,y^{1/2}
 \int_{-1}^{1}\!dz\,\frac{\Phi_N(z)}{\sqrt{1-z^2}}
 \frac{1}{(1-z^2)^2(1-y^2z^2)^3}\nonumber\\
 &\times&\!\!\!\Bigl\{\,1+3y+y^2-(2-y-7y^2+19y^3-3y^4)z^2+
 y^2(6-17y+16y^2)z^4\,\Bigr\}\,.
 \nonumber\\
\end{eqnarray}

\vspace{12pt}

\noindent $\bm{J^{PC}=1^{-+}}$

\begin{equation}
A^{1,\;1,\;1}_{1,\;0;\;1}=
 i\frac{512\sqrt{2}}{3\sqrt{3}}\pi\sqrt{\alpha}\alpha_s
 \frac{f_{{\cal Q}}}{M_{{\cal Q}}}\frac{|R_1'(0)|}{M_G^{3/2}}\,(1-y)
 \int_{-1}^{1}\!dz\,\frac{\Phi_N(z)}{\sqrt{1-z^2}}\,
 \frac{1}{(1-z^2)(1-y^2z^2)}\,,
\end{equation}

\begin{equation}
A^{1,\;1,\;1}_{1,\;1;\;0}=
 i\frac{512\sqrt{2}}{3\sqrt{3}}\pi\sqrt{\alpha}\alpha_s
 \frac{f_{{\cal Q}}}{M_{{\cal Q}}}\frac{|R_1'(0)|}{M_G^{3/2}}\,y^{1/2}(1-y)
 \int_{-1}^{1}\!dz\,\frac{\Phi_N(z)}{\sqrt{1-z^2}}\,
 \frac{1-yz^2}{(1-z^2)(1-y^2z^2)^2}\,.
\end{equation}

\vspace{12pt}

\noindent $\bm{J^{PC}=2^{++}}$

\begin{equation}
A^{0,\;2,\;2}_{1,\;0;\;1}=
 -\frac{256}{9\sqrt{3}}\pi\sqrt{\alpha}\alpha_s
 \frac{f_{{\cal Q}}}{M_{{\cal Q}}}\frac{|R_0(0)|}{M_G^{1/2}}
 \int_{-1}^{1}\!dz\,\frac{\Phi_N(z)}{\sqrt{1-z^2}}\frac{1}{1-y^2z^2}\,,
\end{equation}

\begin{equation}
A^{0,\;2,\;2}_{1,\;1;\;0}=
 -\frac{256}{9}\pi\sqrt{\alpha}\alpha_s
 \frac{f_{{\cal Q}}}{M_{{\cal Q}}}\frac{|R_0(0)|}{M_G^{1/2}}\,y^{1/2}
 \int_{-1}^{1}\!dz\,\frac{\Phi_N(z)}{\sqrt{1-z^2}}\frac{1}{1-y^2z^2}\,,
\end{equation}

\begin{equation}
A^{0,\;2,\;2}_{1,\,2;-1}=
 -\frac{256\sqrt{2}}{9}\pi\sqrt{\alpha}\alpha_s
 \frac{f_{{\cal Q}}}{M_{{\cal Q}}}\frac{|R_0(0)|}{M_G^{1/2}}\, y
 \int_{-1}^{1}\!dz\,\frac{\Phi_N(z)}{\sqrt{1-z^2}}\frac{1}{1-y^2z^2}\,,
\end{equation}

\begin{eqnarray}
A^{2,\;0,\;2}_{1,\;0;\;1}&=&
 \frac{1024\sqrt{10}}{9\sqrt{3}}\pi\sqrt{\alpha}\alpha_s
 \frac{f_{{\cal Q}}}{M_{{\cal Q}}}\frac{|R_2''(0)|}{M_G^{5/2}}\,
 \int_{-1}^{1}\!dz\,\frac{\Phi_N(z)}{\sqrt{1-z^2}}\nonumber\\
 &\times&\frac{(1-y)^2-y(2-10y+10y^2-3y^3)z^2+
 y^2(3-14y+13y^2-4y^3)z^4+y^4z^6}
 {(1-z^2)^2(1-y^2z^2)^3}\,, \nonumber\\
\end{eqnarray}

\begin{equation}
A^{2,\;0,\;2}_{1,\;1;\;0}=
 -\frac{1024\sqrt{10}}{9}\pi\sqrt{\alpha}\alpha_s
 \frac{f_{{\cal Q}}}{M_{{\cal Q}}}\frac{|R_2''(0)|}{M_G^{5/2}}\,y^{1/2}
 \int_{-1}^{1}\!dz\,\frac{\Phi_N(z)}{\sqrt{1-z^2}}
 \frac{(1-6y+6y^2-2y^3)z^2+y^2z^4}{(1-z^2)^2(1-y^2z^2)^2}\,,
\end{equation}

\begin{equation}
A^{2,\;0,\;2}_{1,\,2;-1}=
 \frac{4096\sqrt{5}}{3\sqrt{3}}\pi\sqrt{\alpha}\alpha_s
 \frac{f_{{\cal Q}}}{M_{{\cal Q}}}\frac{|R_2''(0)|}{M_G^{5/2}}\,y\,
 \int_{-1}^{1}\!dz\,\frac{\Phi_N(z)}{\sqrt{1-z^2}}
 \frac{z^2}{(1-z^2)^2(1-y^2z^2)}\,,
\end{equation}

\begin{eqnarray}
A^{2,\;2,\;2}_{1,\;0;\;1}&=&
 \frac{256\sqrt{10}}{9\sqrt{21}}\pi\sqrt{\alpha}\alpha_s
 \frac{f_{{\cal Q}}}{M_{{\cal Q}}}\frac{|R_2''(0)|}{M_G^{5/2}}\,
 \int_{-1}^{1}\!dz\,\frac{\Phi_N(z)}{\sqrt{1-z^2}}
 \frac{1}{(1-z^2)^2(1-y^2z^2)^3}\nonumber\\
 &\times&\!\!\!\Bigl\{\,21-4(1-y)^2+(9-28y-46y^2+28y^3-12y^4)z^2\nonumber\\
 &&+\,y^2(-18+92y-67y^2+28y^3)z^4-7y^4z^6\,\Bigr\}\,,
\end{eqnarray}

\begin{eqnarray}
A^{2,\;2,\;2}_{1,\;1;\;0}&=&
 \frac{256\sqrt{10}}{9\sqrt{7}}\pi\sqrt{\alpha}\alpha_s
 \frac{f_{{\cal Q}}}{M_{{\cal Q}}}\frac{|R_2''(0)|}{M_G^{5/2}}\,y^{1/2}
 \int_{-1}^{1}\!dz\,\frac{\Phi_N(z)}{\sqrt{1-z^2}}
 \frac{1}{(1-z^2)^2(1-y^2z^2)^3}\nonumber\\
 &\times&\!\!\!\Bigl\{\,21-2(1-y)^2+(1-14y-44y^2+14y^3-6y^4)z^2\nonumber\\
 &&+\,y^2(-2+46y-23y^2+14y^3)z^4-7y^4z^6\,\Bigr\}\,,
\end{eqnarray}

\begin{eqnarray}
A^{2,\;2,\;2}_{1,\,2;-1}&=&
 \frac{512\sqrt{5}}{9\sqrt{7}}\pi\sqrt{\alpha}\alpha_s
 \frac{f_{{\cal Q}}}{M_{{\cal Q}}}\frac{|R_2''(0)|}{M_G^{5/2}}\,y\,
 \int_{-1}^{1}\!dz\,\frac{\Phi_N(z)}{\sqrt{1-z^2}}
 \frac{1}{(1-z^2)^2(1-y^2z^2)^3}\nonumber\\
 &\times&\!\!\!\Bigl\{\,21+4(1-y)^2+(5-56y+46y^2-56y^3+12y^4)z^2\nonumber\\
 &&+\,y^2(18-8y+25y^2)z^4-7y^4z^6\,\Bigr\}\,,
\end{eqnarray}

\begin{eqnarray}
A^{4,\;2,\;2}_{1,\;0;\;1}&=& 
 -\frac{256\sqrt{2}}{3\sqrt{21}}\pi
 \sqrt{\alpha}\alpha_{s}\frac{f_{{\cal Q}}}
 {M_{{\cal Q}}}\frac{|R_{_4}^{IV}(0)|}{M_{G}^{9/2}}
 \int_{-1}^{1}\!dz\,\frac{\Phi_N(z)}{\sqrt{1-z^2}}
 \frac{1}{(1-z^2)^4(1-y^2z^2)^5}\nonumber\\
 &\times&\!\!\!\Bigl\{\,13 + 152\,y - 264\,{y^2} + 104\,{y^3} + 16\,{y^4}
 \nonumber \\
 &&+\,2\,\left( 23 + 172\,y - 1090\,{y^2} + 1280\,{y^3} - 244\,{y^4} -
          284\,{y^5} + 80\,{y^6} \right) \,{z^2}\nonumber\\
 &&+\,\left( 5 - 48\,y - 352\,{y^2} + 392\,{y^3} + 5406\,{y^4} -
          10960\,{y^5} + 7592\,{y^6} - 1800\,{y^7} + 80\,{y^8} \right) \,
        {z^4}\nonumber\\
 &&+\, 4\,{y^2}\,\left( -5 + 68\,y + 259\,{y^2} - 1994\,{y^3} +
          2839\,{y^4} - 1408\,{y^5} + 82\,{y^6} + 54\,{y^7} \right) \,
        {z^6}\nonumber\\
 &&+\,{y^4}\,\left( 46 - 1360\,y + 5520\,{y^2} - 6760\,{y^3} +
          3325\,{y^4} - 456\,{y^5} \right) \,{z^8}\nonumber\\
 &&+\,6\,{y^6}\,\left( 34 - 152\,y + 153\,{y^2} - 56\,{y^3} \right) \,
        {z^{10}} + 21\,{y^8}\,{z^{12}}\,\Bigr\}\,,
\end{eqnarray}

\begin{eqnarray}
A^{4,\;2,\;2}_{1,\;1;\;0}&=&
-\frac{256\sqrt{2}}{9\sqrt{7}}\pi
\sqrt{\alpha}\alpha_{s}\frac{f_{{\cal Q}}}
{M_{{\cal Q}}}\frac{|R_{_4}^{IV}(0)|}{M_{G}^{9/2}}\,y^{1/2}\,
\int_{-1}^{1}\!dz\,\frac{\Phi_N(z)}{\sqrt{1-z^2}}
\frac{1}{(1-z^2)^4(1-y^2z^2)^5}\nonumber\\
&\times&\!\!\!\Bigl\{\,-5 + 32\,y + 108\,{y^2} - 40\,{y^3} -
32\,{y^4}\nonumber \\
&&+\,2\,\left( -1 - 86\,y - 142\,{y^2} + 518\,{y^3} -
1114\,{y^4} + 796\,{y^5} - 160\,{y^6} \right) \,{z^2}\nonumber\\
&&+\,\left( -25 - 84\,y - 460\,{y^2} + 7232\,{y^3} - 16926\,{y^4} +
20156\,{y^5} - 11548\,{y^6} + 2760\,{y^7} - 160\,{y^8} \right)
\,{z^4}\nonumber\\
&&+\,4\,{y^2}\,\left( 31 + 311\,y - 2408\,{y^2} +
          4402\,{y^3} - 4199\,{y^4} + 1769\,{y^5} - 167\,{y^6} -
          54\,{y^7} \right) \,z^6\nonumber\\
&&+\,{y^4}\,\left( -686 + 3716\,y - 4884\,{y^2} + 4256\,{y^3} -
          1829\,{y^4} + 372\,{y^5} \right) \,{z^8}\nonumber\\
&&+\,6\,{y^6}\,\left( 2 - 130\,y + 107\,{y^2} - 42\,{y^3} \right) \,
        {z^{10}} + 63\,{y^8}\,{z^{12}}\,\Bigr\}\,,
\end{eqnarray}

\begin{eqnarray}
A^{4,\;2,\;2}_{1,\,2;-1}&=&
-\frac{512}{9\sqrt{7}}\pi
\sqrt{\alpha}\alpha_{s}\frac{f_{{\cal Q}}}
{M_{{\cal Q}}}\frac{|R_{_4}^{IV}(0)|}{M_{G}^{9/2}}\, 
y \int_{-1}^{1}\!dz\,\frac{\Phi_N(z)}{\sqrt{1-z^2}}
\frac{1}{(1-z^2)^4(1-y^2z^2)^5}\nonumber\\
&\times&\!\!\!\Bigl\{\,
143 - 176\,y + 120\,{y^2} - 32\,{y^3} + 8\,{y^4}\nonumber\\
&&+\,2\,\left( 73 - 736\,y + 1246\,{y^2} - 1256\,{y^3} + 700\,{y^4} -
          256\,{y^5} + 40\,{y^6} \right) \,{z^2}\nonumber \\
&&+\,\left( 31 - 144\,y + 1792\,{y^2} - 2720\,{y^3} + 3618\,{y^4} -
          2960\,{y^5} + 1768\,{y^6} - 480\,{y^7} + 40\,{y^8} \right) \,
        {z^4}\nonumber \\
&&+\,4\,{y^2}\,\left( -13 - 284\,y + 413\,{y^2} - 784\,{y^3}
         + 659\,{y^4} - 380\,{y^5} + 74\,{y^6} \right) \,{z^6}\nonumber \\
&&+\,{y^4}\,\left( 362 + 208\,y + 144\,{y^2} + 160\,{y^3} +
          71\,{y^4} \right) \,{z^8}\nonumber\\
&&+\,6\,{y^6}\,\left( -46 + 8\,y - 25\,{y^2} \right) \,{z^{10}} +
       63\,{y^8}\,{z^{12}}\,\Bigr\}\,.
\end{eqnarray}

\vspace{12pt}

\noindent $\bm{J^{PC}=2^{-+}}$

\begin{equation}
A^{1,\;1,\;2}_{1,\;0;\;1}=
 -i\frac{1024\sqrt{2}}{9}\pi\sqrt{\alpha}\alpha_s
 \frac{f_{{\cal Q}}}{M_{{\cal Q}}}\frac{|R_1'(0)|}{M_G^{3/2}}\,y(1-y)^2
 \int_{-1}^{1}\!dz\,\frac{\Phi_N(z)}{\sqrt{1-z^2}}\,
 \frac{z^2}{(1-z^2)(1-y^2z^2)^2}\,,
\end{equation}

\begin{equation}
A^{1,\;1,\;2}_{1,\;1;\;0}=
 -i\frac{512\sqrt{2}}{3\sqrt{3}}\pi\sqrt{\alpha}\alpha_s
 \frac{f_{{\cal Q}}}{M_{{\cal Q}}}\frac{|R_1'(0)|}{M_G^{3/2}}\,y^{3/2}(1-y)^2
 \int_{-1}^{1}\!dz\,\frac{\Phi_N(z)}{\sqrt{1-z^2}}\,
 \frac{z^2}{(1-z^2)(1-y^2z^2)^2}\,,
\end{equation}

\begin{equation}
A^{1,\;1,\;2}_{1,\,2;-1}=0\,, \text{\hspace{11cm}}
\end{equation}

\begin{eqnarray}
A^{3,\;1,\;2}_{1,\;0;\;1}&=&
 i\frac{2048}{9\sqrt{3}}\pi\sqrt{\alpha}\alpha_s
 \frac{f_{{\cal Q}}}{M_{{\cal Q}}}\frac{|R_3'''(0)|}{M_G^{7/2}}\,y(1-y)^2\,
 \int_{-1}^{1}\!dz\,\frac{\Phi_N(z)}{\sqrt{1-z^2}}
 \frac{1}{(1-z^2)^3(1-y^2z^2)^4}\nonumber\\
 &\times&\!\!\!\Bigl\{\,7+(33-80y+26y^2)z^2
 -(2+16y-78y^2+16y^3+9y^4)z^4\nonumber\\
 &&+\,y^2(12-80y+33y^2)z^6+14y^4z^8\,\Bigr\}\,,
\end{eqnarray}

\begin{eqnarray}
A^{3,\;1,\;2}_{1,\;1;\;0}&=&
 i\frac{1024}{27}\pi\sqrt{\alpha}\alpha_s
 \frac{f_{{\cal Q}}}{M_{{\cal Q}}}
 \frac{|R_3'''(0)|}{M_G^{7/2}}\,y^{1/2}(1-y)^2\,
 \int_{-1}^{1}\!dz\,\frac{\Phi_N(z)}{\sqrt{1-z^2}}
 \frac{1}{(1-z^2)^3(1-y^2z^2)^4}\nonumber\\
 &\times&\!\!\!\Bigl\{\,14(1-y)+(28-157y+188y^2-52y^3)z^2-
 y(3-144y+142y^2+10y^3-18y^4)z^4\nonumber\\
 &&-\,y^3(66-20y-11y^2)z^6+21y^5z^8\,\Bigr\}\,,
\end{eqnarray}

\begin{equation}
A^{3,\;1,\;2}_{1,\,2;-1}=
 -i\frac{28672\sqrt{2}}{27}\pi\sqrt{\alpha}\alpha_s
 \frac{f_{{\cal Q}}}{M_{{\cal Q}}}
 \frac{|R_3'''(0)|}{M_G^{7/2}}\,y(1-y)^2\,
 \int_{-1}^{1}\!dz\,\frac{\Phi_N(z)}{\sqrt{1-z^2}}
 \frac{z^2}{(1-z^2)^3(1-y^2z^2)^2}\,.
\end{equation}

\vspace{12pt}

\noindent $\bm{J^{PC}=3^{++}}$

\begin{equation}
A^{2,\;2,\;3}_{1,\;0;\;1}=0 \,,\text{\hspace{11cm}}
\end{equation}

\begin{equation}
A^{2,\;2,\;3}_{1,\;1;\;0}=
 \frac{1024}{9}\pi\sqrt{\alpha}\alpha_s
 \frac{f_{{\cal Q}}}{M_{{\cal Q}}}
 \frac{|R_2''(0)|}{M_G^{5/2}}\,y^{1/2}(1-y)^2
 \int_{-1}^{1}\!dz\,\frac{\Phi_N(z)}{\sqrt{1-z^2}}
 \frac{1+(3-8y+3y^2)z^2+y^2z^4}{(1-z^2)^2(1-y^2z^2)^3}\,,
\end{equation}

\begin{equation}
A^{2,\;2,\;3}_{1,\,2;-1}=
 \frac{1024\sqrt{5}}{9}\pi\sqrt{\alpha}\alpha_s
 \frac{f_{{\cal Q}}}{M_{{\cal Q}}}\frac{|R_2''(0)|}{M_G^{5/2}}\,y(1-y)^2
 \int_{-1}^{1}\!dz\,\frac{\Phi_N(z)}{\sqrt{1-z^2}}
 \frac{1+(3-8y+3y^2)z^2+y^2z^4}{(1-z^2)^2(1-y^2z^2)^3}\,,
\end{equation}

\begin{equation}
A^{4,\;2,\;3}_{1,\;0;\;1}= 
 \frac{10240}{\sqrt{15}}\pi\sqrt{\alpha}\alpha_{s}\frac{f_{{\cal Q}}}
 {M_{{\cal Q}}}\frac{|R_{_4}^{IV}(0)|}{M_{G}^{9/2}}\,y(1-y)^2
 \int_{-1}^{1}\!dz\,\frac{\Phi_N(z)}{\sqrt{1-z^2}}\,
 \frac{1+(3-8y+3y^2)z^2+y^2z^4}{(1-z^2)^4(1-y^2z^2)^3}\,,
\end{equation}

\begin{eqnarray}
A^{4,\;2,\;3}_{1,\;1;\;0}&=&
-\frac{2560\sqrt{2}}{9\sqrt{5}}\pi
\sqrt{\alpha}\alpha_{s}\frac{f_{{\cal Q}}}
{M_{{\cal Q}}}\frac{|R_{_4}^{IV}(0)|}{M_{G}^{9/2}}\,
y^{1/2}(1-y)^2
\int_{-1}^{1}\!dz\,\frac{\Phi_N(z)}{\sqrt{1-z^2}}\,
\frac{1}{(1-z^2)^4(1-y^2z^2)^5}\nonumber\\
&\times&\!\!\!\Bigl\{\,5 - 4\,y - 4\,{y^2} +
\left( 20 - 40\,y - 59\,{y^2} + 128\,{y^3} -
40\,{y^4} \right) \,{z^2}\nonumber\\
&&+\,\left( 7 - 28\,y - 152\,{y^2} + 664\,{y^3} -
 737\,{y^4} + 260\,{y^5} - 20\,{y^6} \right) \,{z^4}\nonumber\\
&&+\,{y^2}\,\left( -25 + 272\,y - 716\,{y^2} + 632\,{y^3} -
 169\,{y^4} \right) \,{z^6}\nonumber\\
&&+\,{y^4}\,\left( -43 + 140\,y - 112\,{y^2} + 24\,{y^3} \right) \,
 {z^8} - 3\,{y^6}\,{z^{10}}\,\Bigr\}\,,
\end{eqnarray}

\begin{eqnarray}
A^{4,\;2,\;3}_{1,\,2;-1}&=&
-\frac{512\sqrt{2}}{9}\pi\sqrt{\alpha}\alpha_{s}\frac{f_{{\cal Q}}}
{M_{{\cal Q}}}\frac{|R_{_4}^{IV}(0)|}{M_{G}^{9/2}}\,y(1-y)^2
\int_{-1}^{1}\!dz\,\frac{\Phi_N(z)}{\sqrt{1-z^2}}\,
\frac{1}{(1-z^2)^4(1-y^2z^2)^5}\nonumber\\
&\times&\!\!\!\Bigl\{\, 53 - 16\,y + 8\,{y^2} +
 \left( 200 - 712\,y + 589\,{y^2} - 352\,{y^3} + 80\,{y^4} \right)
 \,{z^2}\nonumber\\
&&+\,\left( -5 - 280\,y + 1048\,{y^2} - 1136\,{y^3} +
 1063\,{y^4} - 400\,{y^5} + 40\,{y^6} \right) \,{z^4}\nonumber\\
&&+\,{y^2}\,\left( 155 - 592\,y + 664\,{y^2} - 712\,{y^3} +
 215\,{y^4} \right) \,{z^6}\nonumber\\
&&+\,{y^4}\,\left( -7 + 104\,y + 8\,{y^2} \right) \,{z^8} -
 15\,{y^6}\,{z^{10}}\,\Bigr\}\,.
\end{eqnarray}

\vspace{12pt}

\noindent $\bm{J^{PC}=3^{-+}}$

\begin{eqnarray}
A^{3,\;1,\;3}_{1,\;0;\;1}&=&
 i\frac{512\sqrt{7}}{9}\pi\sqrt{\alpha}\alpha_s
 \frac{f_{{\cal Q}}}{M_{{\cal Q}}}
 \frac{|R_3'''(0)|}{M_G^{7/2}}\,(1-y)^2\,
 \int_{-1}^{1}\!dz\,\frac{\Phi_N(z)}{\sqrt{1-z^2}}
 \frac{1}{(1-z^2)^3(1-y^2z^2)^3}\nonumber\\
 &\times&\!\!\!\Bigl\{1-y+(1-11y+11y^2-3y^3)z^2+(3-y)y^2z^4\Bigr\}\,,
\end{eqnarray}

\begin{eqnarray}
A^{3,\;1,\;3}_{1,\;1;\;0}&=&
i\frac{4096\sqrt{14}}{27}\pi\sqrt{\alpha}\alpha_s
\frac{f_{{\cal Q}}}{M_{{\cal Q}}}
 \frac{|R_3'''(0)|}{M_G^{7/2}}\,y^{1/2}(1-y)^2\,
 \int_{-1}^{1}\!dz\,\frac{\Phi_N(z)}{\sqrt{1-z^2}}
 \frac{1}{(1-z^2)^3(1-y^2z^2)^4}\nonumber\\
 &\times&\!\!\!\Bigl\{1-y-(1+14y-16y^2+5y^3)z^2-y(3-24y+20y^2-7y^3)z^4
 -y^3(3-y+2y^2)z^6\Bigr\}\,,
\end{eqnarray}

\begin{equation}
A^{3,\;1,\;3}_{1,\,2;-1}=
 -i\frac{8192\sqrt{70}}{27}\pi\sqrt{\alpha}\alpha_s
 \frac{f_{{\cal Q}}}{M_{{\cal Q}}}
 \frac{|R_3'''(0)|}{M_G^{7/2}}\,y(1-y)^2\,
 \int_{-1}^{1}\!dz\,\frac{\Phi_N(z)}{\sqrt{1-z^2}}
 \frac{z^2}{(1-z^2)^3(1-y^2z^2)^2}\,.
\end{equation}

\vspace{12pt}

\noindent $\bm{J^{PC}=4^{++}}$

\begin{equation}
A^{2,\;2,\;4}_{1,\;0;\;1}=
 \frac{1024\sqrt{2}}{3\sqrt{21}}\pi\sqrt{\alpha}\alpha_s
 \frac{f_{{\cal Q}}}{M_{{\cal Q}}}\frac{|R_2''(0)|}{M_G^{5/2}}\,(1-y)^2
 \int_{-1}^{1}\!dz\,\frac{\Phi_N(z)}{\sqrt{1-z^2}}
 \frac{1+(3-8y+3y^2)z^2+y^2z^4}{(1-z^2)^2(1-y^2z^2)^3}\,,
\end{equation}

\begin{equation}
A^{2,\;2,\;4}_{1,\;1;\;0}=
 \frac{1024\sqrt{5}}{3\sqrt{21}}\pi\sqrt{\alpha}\alpha_s
 \frac{f_{{\cal Q}}}{M_{{\cal Q}}}\frac{|R_2''(0)|}{M_G^{5/2}}\,y^{1/2}(1-y)^2
 \int_{-1}^{1}\!dz\,\frac{\Phi_N(z)}{\sqrt{1-z^2}}
 \frac{1+(3-8y+3y^2)z^2+y^2z^4}{(1-z^2)^2(1-y^2z^2)^3}\,,
\end{equation}

\begin{equation}
A^{2,\;2,\;4}_{1,\,2;-1}=
 \frac{1024\sqrt{5}}{3\sqrt{21}}\pi\sqrt{\alpha}\alpha_s
 \frac{f_{{\cal Q}}}{M_{{\cal Q}}}\frac{|R_2''(0)|}{M_G^{5/2}}\,y(1-y)^2
 \int_{-1}^{1}\!dz\,\frac{\Phi_N(z)}{\sqrt{1-z^2}}
 \frac{1+(3-8y+3y^2)z^2+y^2z^4}{(1-z^2)^2(1-y^2z^2)^3}\,,
\end{equation}

\begin{eqnarray}
A^{4,\;0,\;4}_{1,\;0;\;1}&=&
-\frac{2\sqrt{2}}{\sqrt{3}}\pi\sqrt{\alpha}\alpha_{s}\frac{f_{{\cal Q}}}
{M_{{\cal Q}}}\frac{|R_{_4}^{IV}(0)|}{M_{G}^{9/2}} 
\int_{-1}^{1}\!dz\,\frac{\Phi_N(z)}{\sqrt{1-z^2}}\,
\frac{1}{(1-z^2)^4(1-y^2z^2)^5}\nonumber\\
&\times&\!\!\!\Bigl\{\,(1-y)^2\,
 \left( 391 - 6102\,y + 1721\,{y^2} \right)\nonumber \\
&&+\,\left( 2503 - 23466\,y + 139087\,{y^2} - 301024\,{y^3} +
 279356\,{y^4} - 114996\,{y^5} + 17210\,{y^6} \right) \,{z^2}\nonumber \\
&&+\,y\,\left( -3442 + 67196\,y - 413514\,{y^2} + 929047\,{y^3} -
 915276\,{y^4} + 425324\,{y^5} - 89960\,{y^6} + 8605\,{y^7} \right)
 \,{z^4}\nonumber \\
&&+\,{y^2}\,\left( 625 - 40670\,y + 320322\,{y^2} - 798058\,{y^3} +
 773961\,{y^4} - 321024\,{y^5} + 52012\,{y^6} - 5788\,{y^7} \right)
 \,{z^6}\nonumber \\
&&+\,2\,{y^4}\,\left( 1955 - 46467\,y + 153058\,{y^2} - 141519\,{y^3} +
 42517\,{y^4} + 1096\,{y^5} \right) \,{z^8}\nonumber \\
&&+\, {y^6}\,\left( 20341 - 83242\,y + 74871\,{y^2} - 23940\,{y^3} \right)
 \,{z^{10}} + 2660\,{y^8}\,{z^{12}}\,\Bigr\}\,,
\end{eqnarray}

\begin{eqnarray}
A^{4,\;0,\;4}_{1,\;1;\;0}&=&
\frac{20480}{3\sqrt{15}}\pi\sqrt{\alpha}\alpha_{s}\frac{f_{{\cal Q}}}
{M_{{\cal Q}}}\frac{|R_{_4}^{IV}(0)|}{M_{G}^{9/2}}\,y^{1/2}(1-y)^2
\int_{-1}^{1}\!dz\,\frac{\Phi_N(z)}{\sqrt{1-z^2}}\,
\frac{z^2}{(1-z^2)^4(1-y^2z^2)^4}\nonumber\\
&\times&\!\!\!\Bigl\{\, 3 - 12\,y + 12\,{y^2} - 4\,{y^3} +
 \left( 1 - 20\,y + 62\,{y^2} - 64\,{y^3} + 28\,{y^4} -
 4\,{y^5} \right) \,{z^2}\nonumber \\
&&+\,{y^2}\,\left( 6 - 20\,y + 15\,{y^2} - 4\,{y^3} \right) \,{z^4} +
 {y^4}\,{z^6}\,\Bigr\}\,,
\end{eqnarray}

\begin{eqnarray}
A^{4,\;0,\;4}_{1,\,2;-1}&=& 
-\frac{20480}{3\sqrt{15}}\pi\sqrt{\alpha}\alpha_{s}\frac{f_{{\cal Q}}}
{M_{{\cal Q}}}\frac{|R_{_4}^{IV}(0)|}{M_{G}^{9/2}}\,y(1-y)^2
\int_{-1}^{1}\!dz\,\frac{\Phi_N(z)}{\sqrt{1-z^2}}\,
\frac{z^2\left[\,1 + \left( 3 - 8\,y + 3\,{y^2} \right)
\,{z^2} + {y^2}\,{z^4}\,\right]}
{(1-z^2)^4(1-y^2z^2)^3}\,,
\end{eqnarray}

\begin{eqnarray}
A^{4,\;2,\;4}_{1,\;0;\;1}&=& 
-\frac{5120}{3\sqrt{1155}}\pi\sqrt{\alpha}\alpha_{s}\frac{f_{{\cal Q}}}
{M_{{\cal Q}}}\frac{|R_{_4}^{IV}(0)|}{M_{G}^{9/2}}\,(1-y)^2
\int_{-1}^{1}\!dz\,\frac{\Phi_N(z)}{\sqrt{1-z^2}}\,
\frac{1}{(1-z^2)^4(1-y^2z^2)^5}\nonumber\\
&\times&\!\!\!\Bigl\{\,25 + 16\,y - 8\,{y^2} +
 \left( 96 - 176\,y - 247\,{y^2} + 264\,{y^3} - 80\,{y^4} \right) \,
 {z^2}\nonumber \\
&&+\,\left( 15 - 128\,y - 144\,{y^2} + 1240\,{y^3} -
 1101\,{y^4} + 400\,{y^5} - 40\,{y^6} \right) \,{z^4}\nonumber \\
&&+\,{y^2}\,\left( -25 + 528\,y - 1600\,{y^2} + 1408\,{y^3} -
 597\,{y^4} + 88\,{y^5} \right) \,{z^6}\nonumber \\
&&+\,{y^4}\,\left( -107 + 368\,y - 272\,{y^2} + 88\,{y^3} \right) \,
 {z^8} - 11\,{y^6}\,{z^{10}}\,\Bigr\}\,,
\end{eqnarray}

\begin{eqnarray}
A^{4,\;2,\;4}_{1,\;1;\;0}&=&
-\frac{512\sqrt{2}}{3\sqrt{231}}\pi\sqrt{\alpha}\alpha_{s}\frac{f_{{\cal Q}}}
{M_{{\cal Q}}}\frac{|R_{_4}^{IV}(0)|}{M_{G}^{9/2}}\,y^{1/2}(1-y)^2
\int_{-1}^{1}\!dz\,\frac{\Phi_N(z)}{\sqrt{1-z^2}}\,
\frac{1}{(1-z^2)^4(1-y^2z^2)^5}\nonumber\\
&\times&\!\!\!\Bigl\{\,3\,\left( 51 + 8\,y - 4\,{y^2} \right)  +
\left( 540 - 1452\,y + 9\,{y^2} + 308\,{y^3} - 120\,{y^4} \right)
 \,{z^2}\nonumber \\
&&+\,\left( -5 - 500\,y + 488\,{y^2} + 2564\,{y^3} -
 1877\,{y^4} + 600\,{y^5} - 60\,{y^6} \right) \,{z^4}\nonumber \\
&&+\,{y^2}\,\left( 155 + 528\,y - 3236\,{y^2} + 2508\,{y^3} -
 1165\,{y^4} + 220\,{y^5} \right) \,{z^6}\nonumber \\
&&+\,{y^4}\,\left( -287 + 1124\,y - 672\,{y^2} + 220\,{y^3} \right) \,
 {z^8} - 55\,{y^6}\,{z^{10}}\,\Bigr\}\,,
\end{eqnarray}

\begin{eqnarray}
A^{4,\;2,\;4}_{1,\,2;-1}&=&
-\frac{512\sqrt{2}}{3\sqrt{231}}\pi\sqrt{\alpha}\alpha_{s}\frac{f_{{\cal Q}}}
{M_{{\cal Q}}}\frac{|R_{_4}^{IV}(0)|}{M_{G}^{9/2}}\,y(1-y)^2
\int_{-1}^{1}\!dz\,\frac{\Phi_N(z)}{\sqrt{1-z^2}}\,
\frac{1}{(1-z^2)^4(1-y^2z^2)^5}\nonumber\\
&\times&\!\!\!\Bigl\{\,3\,\left( 79 - 48\,y + 24\,{y^2} \right)  +
 \left( 1160 - 4488\,y + 5061\,{y^2} - 3168\,{y^3} +
 720\,{y^4} \right) \,{z^2}\nonumber \\
&&+\,\left( 195 - 3160\,y + 10712\,{y^2} - 14064\,{y^3} +
 10767\,{y^4} - 3600\,{y^5} + 360\,{y^6} \right) \,{z^4}\nonumber \\
&&+\,{y^2}\,\left( 995 - 4048\,y + 5336\,{y^2} - 4488\,{y^3} +
 1215\,{y^4} \right) \,{z^6}\nonumber \\
&&+\,{y^4}\,\left( 17 + 296\,y + 72\,{y^2} \right) \,{z^8} -
 55\,{y^6}\,{z^{10}}\,\Bigr\}\,.
\end{eqnarray}

\vspace{12pt}

\noindent $\bm{J^{PC}=4^{-+}}$

\begin{eqnarray}
A^{3,\;1,\;4}_{1,\;0;\;1}&=&  
i\frac{16384}{9}\pi\sqrt{\alpha}\alpha_{s}
\frac{f_{{\cal Q}}}
{M_{{\cal Q}}}\frac{|R_{_3}'''(0)|}{M_{G}^{7/2}}\,y(1-y)^4
\int_{-1}^{1}\!dz\,\frac{\Phi_N(z)}{\sqrt{1-z^2}}\,
\frac{z^2\left[\,1 + \left( 1 - 4\,y + {y^2} \right) \,{z^2}
 + {y^2}\,{z^4}\,\right]}
{(1-z^2)^3(1-y^2z^2)^4}\,,
\end{eqnarray}

\begin{eqnarray}
A^{3,\;1,\;4}_{1,\;1;\;0}&=&
i\frac{20480\sqrt{2}}{9\sqrt{5}}\pi\sqrt{\alpha}\alpha_{s}
\frac{f_{{\cal Q}}}
{M_{{\cal Q}}}\frac{|R_{_3}'''(0)|}{M_{G}^{7/2}}\,y^{3/2}(1-y)^4
\int_{-1}^{1}\!dz\,\frac{\Phi_N(z)}{\sqrt{1-z^2}}\,
\frac{z^2\left[\,1 + \left( 1 - 4\,y + {y^2} \right) \,{z^2}
 + {y^2}\,{z^4}\,\right]}
{(1-z^2)^3(1-y^2z^2)^4}\,,
\end{eqnarray}

\begin{eqnarray}
A^{3,\;1,\;4}_{1,\,2;-1} &=& 0\,. \text{\hspace{13.5cm}} 
\end{eqnarray}

\section{Useful analytical integrals}
\label{anint}

In this Appendix we collect several relations useful for the
complete analytical calculation of the amplitudes
$A_{\lambda_\gamma,\lambda_G;\lambda_{{\cal Q}}}$.
An inspection of the results  presented in Appendix \ref{appb}
shows that all the required integrals 
can be cast in the form of Eq.~(\ref{psint}),

\begin{equation}
\Psi_{l,m,n}(y) = 2\int^1_0\,dz\,\frac{\Phi(z)}{\sqrt{1-z^2}}\,
\frac{z^{2m}}{(1-z^2)^l(1-y^2z^2)^n}\,.
\label{psintac}
\end{equation}

We consider separately the two classes of glueball distribution
amplitudes utilized in this paper.

\subsection{Generalized asymptotic distribution amplitude}
\label{acgas}

Using the generalized asymptotic distribution
amplitude $\Phi^{GAS}_L(z)$, defined in Eq.~(\ref{gasy}), 
we easily see that the integrals $\Psi_{l,m,n}(y)$
reduce to

\begin{eqnarray}
 \Psi^{GAS}_{l,m,n}(y)=2N_L\int_0^1\!dz\,\frac{(1-z^2)^{1/2}\,z^{2m}}
 {(1-y^2z^2)^n}
 = N_L{\cal I}_{m,n}(y)\,,
\label{psintgasy}
\end{eqnarray}

\noindent where ${\cal I}_{m,n}(y)$, independent of $L$, is given by

\begin{eqnarray}
{\cal I}_{m,n}(y) &=& 2\int_0^1\!dz\,\frac{(1-z^2)^{1/2}\,z^{2m}}
 {(1-y^2z^2)^n}\nonumber\\
 &=& \frac{\sqrt{\pi}\ \Gamma(m+1/2)}{2\,\Gamma(m+2)}\ 
 _2F_1(m+1/2,n,m+2,y^2)\,,
\label{imny}
\end{eqnarray}

\noindent and $_{2}F_{1}(a,b,c,z)$ is the well known hypergeometrical
function. It is easy to verify that

\begin{equation}
{\cal I}_{m,0}(y) = {\cal I}_{m,n}(y=0) = \frac{\pi}{2^{m+1}}
 \frac{(2m-1)!!}{(m+1)!}\,.
\end{equation}

Two useful recursive relations between the ${\cal I}_{m,n}(y)$
are the following:

\begin{equation}
{\cal I}_{m+1,n+1}(y)=\frac{1}{2ny}\frac{d}{dy}{\cal I}_{m,n}(y)\,,
\qquad n>0\,,
\end{equation}

\begin{equation}
{\cal I}_{m+1,n+1}(y)=\frac{1}{y^2}\Bigl[{\cal I}_{m,n+1}(y)-
 {\cal I}_{m,n}(y)\Bigr]\,.
\end{equation}

{}From these two relations one can easily get as well the following one:

\begin{equation}
{\cal I}_{m,n+1}(y)=\Bigl(\,1+\frac{y}{2n}\,\frac{d}{dy}\,\Bigr)\,
{\cal I}_{m,n}(y)\,,\qquad n>0\,. 
\end{equation}

Eq.~(\ref{imny}) is not very useful for practical calculations.
One can instead utilize the well known transformation formula for
the hypergeometrical functions,

\begin{eqnarray}
_{2}F_1(a,b,c;x) &=& \frac{\Gamma(c)\,\Gamma(-c')}{\Gamma(a')\,\Gamma(b')}
\  _{2}F_1(a,b,c'+1;1-x) \nonumber\\
&+& (1-x)^{-c'}\ \frac{\Gamma(c)\,\Gamma(c')}{\Gamma(a)\,\Gamma(b)}
\ _{2}F_1(a',b',1-c';1-x)\, ,
\label{hyptra}
\end{eqnarray} 

\noindent
where $a'=c-a$, $b'=c-b$, and $c'=a+b-c$,
together with the so-called Gauss recursion formulas,

\begin{eqnarray}
_{2}F_1(a,b,c;x)&=&_{2}F_1(a+1,b-1,c;x)+\frac{a-b+1}{c}\ x
\ _{2}F_1(a+1,b,c+1;x)\,, \\
_{2}F_1(a,b,c;x)&=&\frac{(c-1)(a-b-1)}{(a-1)(c-b-1)}
\ _{2}F_1(a-1,b,c-1;x)+\frac{b(c-a)}{(a-1)(c-b-1)}
\ _{2}F_1(a-1,b+1,c;x)\,,
\label{gauss}
\end{eqnarray}

\noindent
in order to express ${\cal I}_{m,n}(y)$ as a combination of
several hypergeometrical functions of the simple form
$_{2}F_1(r,d,d;x)=(1-x)^{-r}$, with arbitrary real $d$ and integer $r$.

For completeness, we present below all the functions
${\cal I}_{m,n}(y)$ required in the evaluation of the helicity scattering
amplitudes of interest.

\begin{eqnarray}
{\cal I}_{0,1}(y)&=&\frac{\pi}{y^2}(1-\sqrt{1-y^2})\,, \\
{\cal I}_{1,1}(y)&=&\frac{\pi}{y^4}
 (1-\sqrt{1-y^2}-\frac{1}{2}y^2)\,, \\
{\cal I}_{2,1}(y)&=&\frac{\pi}{y^6}
 (1-\sqrt{1-y^2}-\frac{1}{2}y^2-\frac{1}{8}y^4)\,, \\
{\cal I}_{3,1}(y)&=&\frac{\pi}{y^8}
 (1-\sqrt{1-y^2}-\frac{1}{2}y^2-\frac{1}{8}y^4-\frac{1}{16}y^6)\,, \\
{\cal I}_{4,1}(y)&=&\frac{\pi}{y^{10}}
 (1-\sqrt{1-y^2}-\frac{1}{2}y^2-\frac{1}{8}y^4-\frac{1}{16}y^6-
  \frac{5}{128}y^8)\,.
\end{eqnarray}

\begin{eqnarray}
{\cal I}_{0,2}(y)&=&\frac{\pi}{2}\,\frac{1}{\sqrt{1-y^2}}\,, \\
{\cal I}_{1,2}(y)&=&\frac{\pi}{y^4}
 (\frac{2-y^2}{2\sqrt{1-y^2}}-1)\,, \\
{\cal I}_{2,2}(y)&=&\frac{\pi}{2y^6}\,
 (\frac{4-3y^2}{\sqrt{1-y^2}}-4+y^2)\,, \\
{\cal I}_{3,2}(y)&=&\frac{\pi}{2y^8}\,
 (\frac{6-5y^2}{\sqrt{1-y^2}}-6+2y^2+\frac{1}{4}y^4)\,, \\
{\cal I}_{4,2}(y)&=&\frac{\pi}{2y^{10}}\,
 (\frac{8-7y^2}{\sqrt{1-y^2}}-8+3y^2+\frac{1}{2}y^4+\frac{1}{8}y^6)\,.
\end{eqnarray}

\begin{eqnarray}
{\cal I}_{0,3}(y)&=&\frac{\pi}{8}\,
 \frac{4-3y^2}{(1-y^2)^{3/2}}\,, \\
{\cal I}_{1,3}(y)&=&\frac{\pi}{8}\,\frac{1}{(1-y^2)^{3/2}}\,, \\
{\cal I}_{2,3}(y)&=&\frac{\pi}{y^6}\,
 (1-\frac{8-12y^2+3y^4}{8(1-y^2)^{3/2}})\,, \\
{\cal I}_{3,3}(y)&=&\frac{\pi}{2y^8}\,
 (6-y^2-\frac{24-40y^2+15y^4}{4(1-y^2)^{3/2}})\,, \\
{\cal I}_{4,3}(y)&=&\frac{\pi}{8y^{10}}\,
 (48-12y^2-y^4-\frac{48-84y^2+35y^4}{(1-y^2)^{3/2}})\,.
\end{eqnarray}

\begin{eqnarray}
{\cal I}_{0,4}(y)&=&\frac{\pi}{16}\,
\frac{8-12y^2+5y^4}{(1-y^2)^{5/2}}\,, \\
{\cal I}_{1,4}(y)&=&\frac{\pi}{16}\,
\frac{2-y^2}{(1-y^2)^{5/2}}\,, \\
{\cal I}_{2,4}(y)&=&\frac{\pi}{16}\,
\frac{1}{(1-y^2)^{5/2}}\,, \\
{\cal I}_{3,4}(y)&=&\frac{\pi}{y^8}\,
(\,\frac{16-40y^2+30y^4-5y^6}{16(1-y^2)^{5/2}}-1\,)\,, \\
{\cal I}_{4,4}(y)&=&\frac{\pi}{2y^{10}}\,
(\,\frac{64-168y^2+140y^4-35y^6}{8(1-y^2)^{5/2}}-8+y^2\,)\,.
\end{eqnarray}

\begin{eqnarray}
{\cal I}_{0,5}(y)&=&\frac{\pi}{128}\,
\frac{64-144y^2+120y^4-35y^6}{(1-y^2)^{7/2}}\,, \\
{\cal I}_{1,5}(y)&=&\frac{\pi}{128}\,
\frac{16-16y^2+5y^4}{(1-y^2)^{7/2}}\,, \\
{\cal I}_{2,5}(y)&=&\frac{\pi}{128}\,
\frac{8-3y^2}{(1-y^2)^{7/2}}\,, \\
{\cal I}_{3,5}(y)&=&\frac{\pi}{128}\,
\frac{5}{(1-y^2)^{7/2}}\,, \\
{\cal I}_{4,5}(y)&=&\frac{\pi}{y^{10}}\,
(\,1-\frac{128-448y^2+560y^4-280y^6+35y^8}
{128(1-y^2)^{7/2}}\,)\,, \\
{\cal I}_{5,5}(y)&=&\frac{\pi}{2y^{12}}\,
(\,10-y^2-\frac{640-2304y^2+3024y^4-1680y^6+315y^8}
{64(1-y^2)^{7/2}}\,)\,, \\
{\cal I}_{6,5}(y)&=&\frac{\pi}{8y^{14}}\,
(\,120-20y^2-y^4-\frac{1920-7040y^2+9504y^4-5544y^6+1155y^8}
{16(1-y^2)^{7/2}}\,)\,.
\end{eqnarray}

\subsection{The modified asymptotic distribution amplitude}
\label{acmas} 

Using the modified asymptotic distribution amplitude $\Phi^{AS}(z)$,
defined in Eq.~(\ref{masy}), the integrals $\Psi_{L,m,n}(y)$ take the form

\begin{equation}
\Psi_{l,m,n}(y,u)=2N_u\,\int^{1}_{0}\,dz\,
\frac{\exp[-u/(1-z^2)]\,z^{2m}}{(1-z^2)^{l-1/2}(1-y^2z^2)^{n}}\,,
\label{psintmasy}
\end{equation}

\noindent
with the normalization factor

\begin{equation}
N_u=\frac{3}{2}\, \exp(u/2)\, \left[\,u^2
K_{0}(u/2)+u(1-u)K_{1}(u/2)\,\right]^{-1}\,.
\label{nuac}
\end{equation}

Let us now consider the integrals

\begin{equation}
{\cal J}_{l,m,n}(y,u)=2\,\int^{1}_{0}\,dz\,
\frac{\exp[-u/(1-z^2)]\,z^{2m}}{(1-z^2)^{l-1/2}(1-y^2z^2)^{n}}\,.
\label{jlmn}
\end{equation}

It is easy to verify that the following recurrence relations hold:

\begin{eqnarray}
{\cal J}_{l,m+1,n+1}(y,u)&=&\frac{1}{y^2}\,
\left(\,{\cal J}_{l,m,n+1}(y,u)-{\cal J}_{l,m,n}(y,u)\,\right)\,,\nonumber\\
{\cal J}_{l+1,m+1,n}(y,u)&=&{\cal J}_{l+1,m,n}(y,u)-
{\cal J}_{l,m,n}(y,u)\,,\nonumber\\
{\cal J}_{l,m+1,n+1}(y,u)&=&\frac{1}{2ny}\,\frac{d}{dy}
{\cal J}_{l,m,n}(y,u)\,,\qquad n>0\,, \label{jrec} \\
{\cal J}_{l+1,m,n}(y,u)&=&-\frac{d}{du}{\cal J}_{l,m,n}(y,u)\,,\nonumber\\
{\cal J}_{l,m,n+1}(y,u)&=&\left(\,1+\frac{y}{2n}\,
\frac{d}{dy}\,\right)\,{\cal J}_{l,m,n}(y,u)\,,\qquad n>0\,.\nonumber
\end{eqnarray}

By defining $t+1=1/(1-z^2)$ and $\gamma=1/(1-y^2)$, the integrals
${\cal J}_{l,m,n}(y,u)$ can be cast in the form

\begin{equation}
{\cal J}_{l,m,n}(y,u)=e^{-u}\,\gamma^{n}\,\int^{\,\infty}_{0}\,
dt\,t^{m-1/2}\,(t+1)^{l-m+n-2}\,(t+\gamma)^{-n}\,e^{-ut}\,.
\label{jtgamma}
\end{equation}

Solving these integrals is not immediate. However, one can show
that, by performing a certain number of appropriate integrations
and differentiations
with respect to $u$ (the order and the number of these operations
being related to the values of $l$, $m$, $n$) and using as
boundary conditions the simpler integrals ${\cal J}_{l,m,n}(y,u=0)$, 
they can be related to an integral representation of the well known
error function, $\text{Erf}(x)$.

Without entering into more details of this conceptually simple but
algebraically cumbersome procedure, we present below a minimal set of
three integrals of the family ${\cal J}_{l,m,n}(y,u)$ that,
using the recurrence relations (\ref{jrec}), generate all other
integrals required by the calculation of our scattering amplitudes.
Since the expressions found are in some cases quite lengthy and
involved, we do not present all of them explicitly.

\begin{eqnarray}
{\cal J}_{0,0,0}(y,u)&=&\sqrt{\pi u}\,e^{-u}-\pi(u-1/2)\,
\left[\,1-{\rm Erf}(\sqrt{u})\,\right]\,,\\
{\cal J}_{0,0,1}(y,u)&=&\frac{\pi}{y^2}\,\left\{\,1-{\rm Erf}
(\sqrt{u})-\sqrt{1-y^2}\,\exp\left[\,uy^2/(1-y^2)\,\right]\,
\left[\,1-{\rm Erf}\left(\sqrt{u/(1-y^2)}\right)\,\right]\,\right\}\,,\\
{\cal J}_{0,2,1}(y,u)&=&\frac{\pi}{8y^6}\,e^{-u}\,\Bigl\{\,
-2\sqrt{u}\,y^2\,(4+y^2+2uy^4)
+\left[\,8+(8u-4)y^2+(4u^2+4u-1)y^4\,\right]\sqrt{\pi}\,e^{u}\,
\left[\,1-{\rm Erf}(\sqrt{u})\,\right]\nonumber\\
&-&8\sqrt{\pi(1-y^2)}\,\exp[u/(1-y^2)]\,
\left[\,1-{\rm Erf}\left(\sqrt{u/(1-y^2)}\right)\,\right]\,\Bigr\}\,.
\label{jset3}
\end{eqnarray}


\newpage 

\begin{table}
\caption{For each $|J^{PC}\rangle$ two-gluon glueball state, up to $J=4$,
 the contributing $|L,S\rangle$ states with $L \leq 4$ are shown.}
\begin{ruledtabular}
\begin{tabular}{rl}
\noalign{\vspace{3pt}}
 $|J^{PC}\rangle$ & $|L,S\rangle$ \\
\noalign{\vspace{3pt}}
\colrule
\noalign{\vspace{3pt}}
 $0^{++}$ & $|L=0,S=0\rangle$; $|L=2,S=2\rangle$ \\
 $0^{-+}$ & $|L=1,S=1\rangle$ \\
 $1^{++}$ & $|L=2,S=2\rangle$ \\
 $1^{-+}$ & $|L=1,S=1\rangle$ \\
 $2^{++}$ & $|L=0,S=2\rangle$; $|L=2,S=0\rangle$;
            $|L=2,S=2\rangle$; $|L=4,S=2\rangle$ \\
 $2^{-+}$ & $|L=1,S=1\rangle$; $|L=3,S=1\rangle$ \\
 $3^{++}$ & $|L=2,S=2\rangle$; $|L=4,S=2\rangle$ \\
 $3^{-+}$ & $|L=3,S=1\rangle$ \\
 $4^{++}$ & $|L=2,S=2\rangle$; $|L=4,S=0\rangle$;
            $|L=4,S=2\rangle$ \\
 $4^{-+}$ & $|L=3,S=1\rangle$ \\
\noalign{\vspace{3pt}}
\end{tabular}
\end{ruledtabular}
\label{jls}
\end{table}

\begin{table}
\caption{Numerical values of the ratio $R=\Gamma(\Upsilon\to\gamma\,G)/
\Gamma(J/\psi\to\gamma\,G)$ for the glueball candidates considered
in this paper; for each case, different $|L,S\rangle$ contributions
(up to $L=4$) and three possible choices for the glueball DA (see text)
are considered; comparison with experimental bounds, when available, is also
shown.}
\begin{ruledtabular}
\begin{tabular}{cccrrrr}
\noalign{\vspace{3pt}}
$|\,J^{PC}\,\rangle$ & $M_{G}$ (GeV) & $|\,L\;S\,\rangle$ &
 \multicolumn{3}{c}{$R$} &\multicolumn{1}{c}{$R^{\mbox{\rm\small exp}}$}
 \\
\noalign{\vspace{3pt}}
\colrule
\noalign{\vspace{3pt}}
 & & &  \multicolumn{1}{c}{$\;$GAS$\;$} & \multicolumn{2}{c}{ASI} & \\
\noalign{\vspace{3pt}}
\cline{4-6}
\noalign{\vspace{3pt}}
 & & & & \multicolumn{1}{c}{$\;b=0.5$ GeV$\;$} &
 \multicolumn{1}{c}{$\;b=1.0$ GeV$\;$} & \\
\noalign{\vspace{3pt}}
\cline{5-6}
\noalign{\vspace{3pt}}
 $0^{++}$  &  1.71  &  $0\;\;0$  &  $2.1\,10^{-2}$  &  $2.2\,10^{-2}$  &
 $2.2\,10^{-2}$     &  $< 3.4\,10^{-1}$ \\
\noalign{\vspace{1pt}}
           &        &  $2\;\;2$  &  $3.1\,10^{-3}$  &  $2.1\,10^{-3}$  &
 $2.8\,10^{-3}$     &                   \\
\noalign{\vspace{3pt}}
 $0^{-+}$  &  2.14  &  $1\;\;1$  &  $3.8\,10^{-4}$  &  $3.4\,10^{-4}$  &
 $3.5\,10^{-4}$     &                   \\
\noalign{\vspace{3pt}}
 $1^{++}$  &  2.34  &  $2\;\;2$  &  $2.8\,10^{-4}$  &  $4.3\,10^{-4}$  &
 $3.6\,10^{-4}$     &                   \\
\noalign{\vspace{3pt}}
 $2^{++}$  &  2.22  &  $0\;\;2$  &  $7.7\,10^{-3}$  &  $8.8\,10^{-3}$  &
 $8.5\,10^{-3}$     &  $< 2.9\,10^{-1}$ \\
\noalign{\vspace{1pt}}
           &        &  $2\;\;0$  &  $1.3\,10^{-2}$  &  $2.2\,10^{-1}$  &
 $4.5\,10^{-2}$     &                   \\
\noalign{\vspace{1pt}}
           &        &  $2\;\;2$  &  $7.6\,10^{-3}$  &  $6.7\,10^{-3}$  &
 $7.1\,10^{-3}$     &                   \\
\noalign{\vspace{1pt}}
           &        &  $4\;\;2$  &  $2.1\,10^{-2}$  &  $7.7\,10^{-3}$  &
 $1.8\,10^{-2}$     &                   \\
\noalign{\vspace{3pt}}
 $2^{-+}$  &  2.04  &  $1\;\;1$  &  $1.3\,10^{-3}$  &  $1.8\,10^{-3}$ &
 $1.6\,10^{-3}$     &                   \\
\noalign{\vspace{1pt}}
           &        &  $3\;\;1$  &  $1.5\,10^{-2}$  &  $1.6\,10^{-2}$  &
 $1.7\,10^{-2}$     &                   \\
\noalign{\vspace{3pt}}
 $3^{++}$  &  2.00  &  $2\;\;2$  &  $1.4\,10^{-2}$  &  $9.7\,10^{-3}$  &
 $1.2\,10^{-2}$     &                   \\
\noalign{\vspace{1pt}}
           &        &  $4\;\;2$  &  $1.0\,10^{-2}$  &  $5.6\,10^{-3}$  &
 $1.0\,10^{-2}$     &                   \\
\noalign{\vspace{3pt}}
 $4^{++}$  &  2.22  &  $2\;\;2$  &  $3.2\,10^{-1}$  &  $2.0\,10^{-1}$  &
 $2.7\,10^{-1}$     &  $<2.9\,10^{-1}$    \\
\noalign{\vspace{1pt}}
           &        &  $4\;\;0$  &  $5.5\,10^{-1}$  &  $2.8\,10^{-3}$  &
 $1.4\,10^{-1}$     &                   \\
\noalign{\vspace{1pt}}
           &        &  $4\;\;2$  &  $3.1\,10^{-1}$  &  $1.5\,10^{-1}$  &
 $2.8\,10^{-1}$     &                   \\
\noalign{\vspace{3pt}}
\end{tabular}
\end{ruledtabular}
\label{rtab}
\end{table}

\newpage
\vspace*{-80pt}

\newpage

\vspace*{2.2cm}

\newpage

\begin{center}
\begin{figure}[t]
\epsfig{file=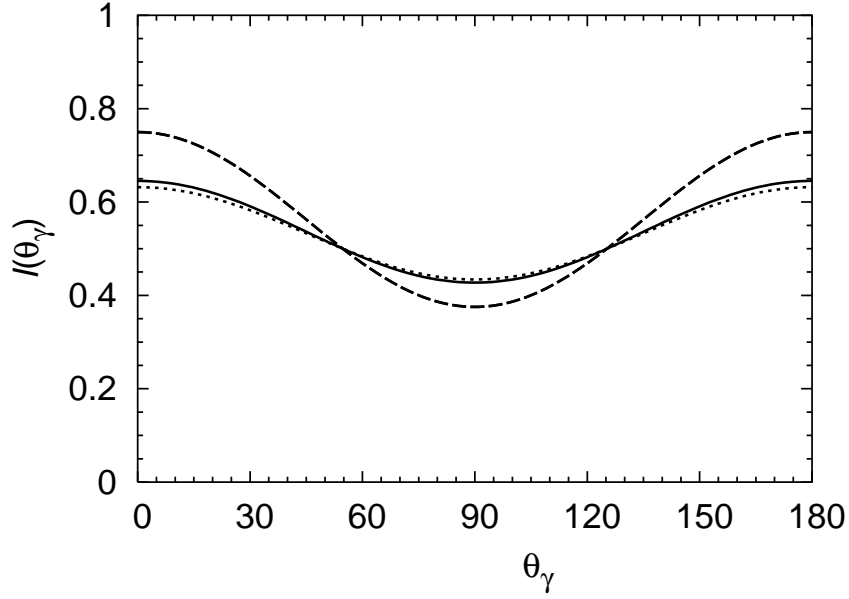,width=11.6truecm}
\caption{Normalized photon angular distribution, $I(\theta_\gamma)=
(1/\Gamma)\,d\,\Gamma/d(\cos\theta_\gamma)$, for the $\Upsilon\to\gamma
\,G(2^{++})$ decay and different $|L,S\,\rangle$ glueball states, using
the asymptotic DA (ASY DA), with $b=0.5$ GeV:
$L=0$, $S=2$ (solid); $L=2$, $S=0$ (dashed); $L=2$, $S=2$ (dotted).
Notice that the $L=4$, $S=2$ case (which is not shown, for clarity)
is almost indistinguishable from the $L=2$, $S=0$ case.
Moreover, notice that for the $L=0$, $S=2$ case $I(\theta_\gamma)$
is independent of the glueball DA; for all other cases, use of the ASY DA
with $b=1.0$ GeV, or of the GAS DA, leads to very similar results.}
\label{u2ppa}
\end{figure}
\begin{figure}[t]
\epsfig{file=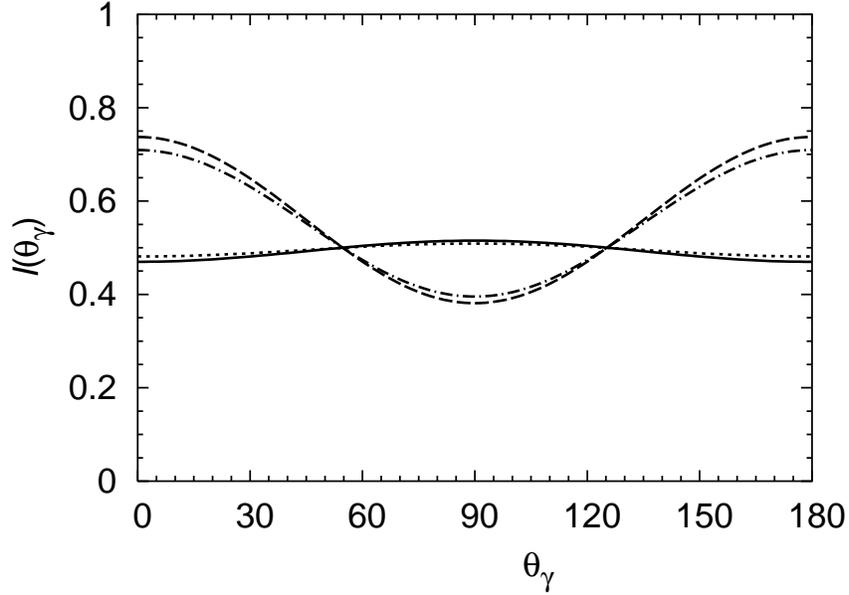,width=11.6truecm}
\caption{Normalized photon angular distribution, $I(\theta_\gamma)=
(1/\Gamma)\,d\,\Gamma/d(\cos\theta_\gamma)$, for the $J/\psi\to\gamma
\,G(2^{++})$ decay and different $|L,S\,\rangle$ glueball states, using
the asymptotic DA (ASY DA), with $b=0.5$ GeV:
$L=0$, $S=2$ (solid); $L=2$, $S=0$ (dashed); $L=2$, $S=2$ (dotted);
$L=4$, $S=2$ (dot-dashed). Notice that for the $L=0$, $S=2$ case
$I(\theta_\gamma)$ is independent of the glueball DA; for all other cases,
use of the ASY DA with $b=1.0$ GeV, or of the GAS DA,
leads to very similar results.}
\label{j2ppa}
\vspace{-20pt}
\end{figure}
\end{center}

\newpage


\newpage


\newpage


\begin{thebibliography}{}
\bibitem{exp}    See e.g.:
                 C.A. Meyer,
                 Proceedings of the Workshop on Gluonic Excitations,
                 Newport News, Virginia, 2003, AIP Conf. Proc. {\bf                                 698}, 554 (2004);
                 Z.-P. Zheng,
                 Proceedings of the International Symposium on
                 Hadron spectroscopy, chiral symmetry and relativistic
                 description of bound systems, Tokyo, 2003, edited by
                 K. Takamatsu {\it et al.}, p. 171;
                 U. Wiedner,
                 in JHW2002, Proceedings of the Twenty-sixth Johns Hopkins                          Workshop, Heidelberg, Germany, edited by G. Domokos
                 {\it et al.}, JHEP Proceedings, PrHEP JHW2002;
                 K.K. Seth,
                 Nucl. Phys. B (Proc. Suppl.) {\bf 96}, 205 (2001).
\bibitem{zou}    Bing-Song Zou,
                 Nucl. Phys. {\bf A655}, 41c (1999).
\bibitem{morn}   C.J. Morningstar and M. Peardon,
                 Phys. Rev. {\bf D60}, 034509 (1999).
\bibitem{wein}   A. Vaccarino and D. Weingarten,
                 Phys. Rev. {\bf D60}, 114501 (1999).
\bibitem{carl}   C.E. Carlson, T.H. Hansson, and C. Peterson,
                 Phys. Rev. {\bf D27}, 1556 (1983).
\bibitem{isgu}   N. Isgur and J. Paton,
                 Phys. Rev. {\bf D31}, 2910 (1985).
\bibitem{nari}   S. Narison,
                 Nucl. Phys. B (Proc. Suppl.) {\bf 96}, 244 (2001),
                 and references therein;
                 S. Narison, {\it QCD Spectral Sum Rules},
                 Lecture Notes in Physics, {\bf Vol. 26}
                 (World Scientific, Singapore, 1989).
\bibitem{chan}   M.S. Chanowitz, Proceedings of the 6th International
                 Workshop on Photon-Photon Collisions, edited by R.L. Lander
                 (World Scientific, Singapore, 1985).
\bibitem{pdg}    The Review of Particle Physics,
                 K. Hagiwara {\it et al.},  
                 Phys. Rev. {\bf D66}, 010001 (2002).
\bibitem{Kada}   H.E. Kada, P. Kessler, and J. Parisi,
                 Phys. Rev. {\bf D39}, 2657 (1989).   
\bibitem{Houra}  L. Houra-Yaou, P. Kessler, and J. Parisi,
                 Phys. Rev. {\bf D45}, 794 (1992).
\bibitem{Kessl}  F. Murgia, P. Kessler, and J. Parisi,
                 Z. Phys {\bf C71}, 483 (1996).  
\bibitem{Murgia} L. Houra-Yaou, P. Kessler, J. Parisi,
                 F. Murgia, and  J. Hansson,
                 Z. Phys. {\bf C76}, 537 (1997).
\bibitem{Brod}   For an extensive review and
                 further references, see e.g. S.J. Brodsky,
                 G.P. Lepage, in \textit{Perturbative Quantum
                 Chromodynamics}, edited by A.H. Mueller (World Scientific,
                 Singapore, 1989).
\bibitem{nrqcd}  See e.g. G.T. Bodwin, hep-ph/0312173, and references
                 therein.
\bibitem{kroll}  P. Kroll, Nucl. Phys. B (Proc. Suppl.) {\bf 64}, 456 (1998),
                 and references therein.
\bibitem{corn}   J.M. Cornwall and A. Soni,
                 Phys. Lett. {\bf B120}, 431, (1983).
\bibitem{dzie}   Z. Dziembowski and L. Mankiewicz,
                 Phys. Rev. Lett. {\bf 58}, 2175 (1987).
\bibitem{ji}     C.-R. Ji, P.L. Chung, and S.R. Cotanch,
                 Phys. Rev. {\bf D45}, 4214 (1992).
\bibitem{huan}   T. Huang, B.-Q. Ma, and Q.-X. Shen,
                 Phys. Rev. {\bf D49} 1490 (1994). 
\end{thebibliography}
\end{document}